\documentstyle[twocolumn,aps,pre,eqsecnum,epsf]{revtex}
\begin{document}
\title {
{\rm
PHYSICAL REVIEW E, submitted \hfill Preprint WIS-94/21/Apr-PH
} \\
{}~~\\
Anomalous Scaling and Fusion Rules in  Hydrodynamic Turbulence}
\author{ Vladimir V. Lebedev$^{*}$ and  Victor S. L'vov$^{**}$}
\address{Department of Physics, Weizmann Institute
of Science, Rehovot, 76100, Israel \ \ \ \ {\rm and} \\
$^{*}$Landau Institute for
Theor. Phys., Ac. of Sci. of Russia,
117940, GSP-1, Moscow, Kosygina 2, Russia\\
$^{**}$Institute of Automation and Electrometry, Ac. of Sci. of
Russia, 630090, Novosibirsk, Russia}
\maketitle
\begin{abstract}
It is shown that the statistical properties of fully developed
hydrodynamic turbulence are characterized by an infinite set of
independent anomalous exponents which describes  the scaling behavior of
hydrodynamic fields constructed from the second and larger powers of the
velocity derivatives.  The energy dissipation field $\varepsilon(t,{\bf
r})$ and the square of the vorticity are the simplest examples of such
fields.  A physical mechanism responsible for anomalous scaling is
discovered and investigated.  We call this mechanism the {\it telescopic
multi-step eddy interaction}.  The essence of this mechanism is the
existence of a very large number $(R/\eta)^{\Delta_j}\gg 1$ of channels of
interaction of large eddies of scale $R$ in the inertial interval with
eddies of viscous scale $\eta$ via a set of eddies of all intermediate
scales between $R$ and $\eta$.  The description of this mechanism in the
consistent analytical theory of turbulence based on the Navier Stokes
equation in the quasi Lagrangian representation is presented.  In the
diagrammatic expansion of the correlation function of the energy
dissipation field $K_ {\varepsilon\varepsilon}(R)$, we have found an
infinite series of logarithmically diverging diagrams. Their summation
leads to a renormalization of the normal Kolmogorov-41 dimensions.  For a
description of the scaling of various hydrodynamic fields an infinite set
of primary fields $O_n$ with independent scaling exponents $\Delta_n$ was
introduced.  We have proposed a symmetry classification of the fields
$O_n$ enabling one to predict relations between scaling the behavior of
different correlation functions. For instance the principal contributions
to the irreducible correlation functions of all scalar fields constructed
from velocity derivatives possess the same scaling behavior with a
so-called ``intermittency exponent" $\mu$.  Further we formulate
restrictions imposed on the structure of correlation functions due to the
incompressibility condition, e.g., the simultaneous correlation function
$\langle\varepsilon{\bf v}\rangle$ (where $\varepsilon$ is the energy
dissipation rate) is equal to zero.  Experimental test for the conformal
symmetry of the turbulent correlation functions are proposed. It is
demonstrated that the anomalous scaling behavior should be revealed in the
asymptotic behavior of correlations function of velocity differences.  A
way to obtain the anomalous exponents from experiments is described.
\end{abstract} 
\section*{Introduction.}
\label{sec:intro}
Problems related to statistical properties of fully developed hydrodynamic
turbulence have been intensively attacked during the last century.
Nevertheless no consensus has yet been achieved on the behavior of the
correlation functions of the turbulent velocity.  Investigators agree that
they should be treated in terms of a scaling behavior but the character of
this scaling is still under discussion.  An analytical  theory of scaling
behavior of hydrodynamic fields constructed from the velocity derivatives
is presented in this paper.  We begin with a brief overview of the history
of the problem in order to introduce the notation and to recall the
underlying ideas.

In 1883 Osborne Reynolds suggested the necessity of a statistical
description of  turbulence at high Reynolds number ${\rm Re}=VL/\nu$,
where $V$ is a characteristic velocity, $L$ is a characteristic length
scale, and $\nu$ is the kinematic molecular viscosity. Note that a typical
value of ${\rm Re}$ for water flow in rivers and for air flow in the
atmosphere is about $10^7 - 10^9$ which is tremendously larger then the
critical value ${\rm Re}_{\rm cr}\sim 10^2$  at which a laminar flow loses
its stability.

The modern concept of hydrodynamic turbulence originates from the
Richardson cascade model (1922)\cite{R922} in which turbulent motions are
produced because of the instability of the laminar flow around the
streamlined body, the geometry of which determines the scale of the
instability $L$.  The eddies with the characteristic scale $L$ which
appear due to the instability are also unstable in their turn and produce
as a result eddies of smaller sizes, which are unstable again and produce
eddies of even smaller sizes  and so on and so forth.  This process
continues until the ``current" Reynolds number dependent on the scale
reaches the critical value ${\rm Re}_{\rm cr}$.  Eddies of smaller sizes
are stable, and their energy dissipates into heat due to viscosity.  In
such a way at ${\rm Re}\gg {\rm Re}_{\rm cr}$  {\it fully developed
turbulence} arises in which there are turbulent motions at all scales
from $L$ down to some ``viscous" scale $\eta\ll L$.  We will refer to
these motions at the scale $r$ as $r$-eddies.

Scaling behavior of velocity differences in fully developed turbulence was
predicted by Kolmogorov and Obukhov more than 50 years ago in their
celebrated papers\cite{K941,O941}. They assumed that the energy which is
pumped on the largest scales $\sim L$ is subsequently transferred from
larger eddies to smaller ones and that the statistics of the energy
pumping is lost, except for the injection rate because it equals (in the
stationary case) the value of the energy flux through each scale.  In such
a way the properties of turbulence in the {\it inertial subrange of
scales} $r$, $L \gg r \gg \eta$, are suggested to be universal and to be
independent both of details of the excitation of the turbulence and of
boundary conditions.  In particular it was assumed that the statistics of
fine scale turbulence (at $r\ll L$) is homogeneous and isotropic. Clearly,
the value of the energy flux is equal to the mean value $\bar\varepsilon =
\langle \varepsilon (t,{\bf r}) \rangle$ of the energy dissipation rate
$\varepsilon$ per unit mass occurring at small scales.

Thus according to the Kolmogorov-Obukhov (1941) picture of fully
developed, homogeneous, isotropic turbulence of an incompressible fluid
(hereafter KO-41) there is only one relevant parameter $\bar\varepsilon$
in the inertial subrange of scales which is the mean value of the energy
dissipation rate.  This means that in KO-41 a simultaneous correlation
function of the velocity (and its derivatives) on a scale $r$ in the
inertial subrange is determined only by $\bar\varepsilon$ and by $r$
itself and is independent both of the outer scale $L$ and of the viscous
scale $\eta$. This allows one to evaluate easily all correlation functions
of turbulent fields in the inertial subrange with the help of dimensional
reasoning (see for example textbooks\cite{LL59,MY75}). Some essential
details  are reproduced below, in Section \ref{sec:kolob}.

However the situation in the theory of turbulence is not as simple as it
appear at  first glance. In the 40's Landau\cite{LL59} indicated that the
above KO-41 picture is not so obvious because of the intermittency
inherent in turbulence. This means that there are relatively calm periods
which are interrupted irregularly by strong turbulent bursts either in
space or in time. Experimental evidence of intermittency was given by
Batchelor and Townsend in 1949\cite{BT49} and then by Kuo and Corrsin in
1971, 1972\cite{KC71,KC72}.  Intermittency leads to strong fluctuations of
the energy dissipation rate $\varepsilon$, thus one may find amongst the
 parameters relevant for the description of turbulent motions of scales
$r$ in the inertial subrange along with  the mean value of the energy flux
$\bar\varepsilon$, the dispersion of the energy flux over the scale $r$.
This dispersion may depend on the parameter $L/r$ characterizing the
number of successive crushings of $r$-eddies required to reduce the scale
from $L$ to $r$.

A number of schemes were proposed to take intermittency into account.  The
first attempts belong to Kolmogorov\cite{K962} and Obukhov\cite{O962} who
proposed the so-called {\sl log-normal model of intermittency}. After them
Novikov and Steward\cite{NS64}, Yaglom\cite{Y966}, Mandelbrot\cite{74,76},
Frisch, Sulem and Nelkin\cite{FSN8} and many others have proposed various
modifications of the KO-41 picture. These attempts to construct
phenomenological models are very important as a method to represent
experimental data and interesting from the theoretical viewpoint.  However
all of them suffer from a lack of connection to the Navier-Stokes equation
which  describes the real fluid dynamics.

Various closure procedures have been also constructed, see for example
Obukhov 1941\cite{Ob41}, Heisenberg 1948\cite{H948}, Kraichnan 1965
\cite{K965}, etc.; for a review see, i.e., \cite{O970}. After the
exclusion  of non-physical approximations one can claim that all these
theories give KO-41 scaling of velocity differences.  However all such
closures are uncontrolled approximations since there is no small parameter
in the theory of turbulence. Therefore such approximations may be
considered as a tool to evaluate some dimensionless constants, but not as
a proof of KO-41.  Indeed, one can say that intermittency corrections
possibly stem from some spatial structures of turbulent flow, like
filaments, etc.  The existence of such structures is reflected in the high
moments of velocity differences and therefore cannot be described within
the framework of any closure procedures. Attempts to reach this phenomenon
may be done in a systematic theory of turbulence without any truncations.

The first attempt to formulate a systematic statistical description of
turbulence directly from the Navier-Stokes equation  was made by
Wyld\cite{W961}. Another derivation of the very same diagrammatic approach
was suggested by Martin, Siggia, Rose\cite{MSR3} in the framework of a
functional integration approach.  Unfortunately these techniques lead to
the representation of the correlation functions in the form of an infinite
series and any truncation of these series breaks the Galilean symmetry of
the problem.  It leads to enormous technical difficulties.  Formally these
arise from the infrared behavior of integrals which become more and more
divergent with increasing order in the perturbation series.  The physical
reason for this is the sweeping effect of eddies of given scale in the
inertial interval by the velocity field of all eddies of larger scales.
In order to eliminate this sweeping effect from the theory Kraichnan
formulated\cite{K977} a perturbation expansion in the terms of the
Lagrangian velocity.  Resulting  Kraichnan's ``Lagrangian-history"
renormalized expansion is a systematic perturbation procedure which
possesses Galilean symmetry order by order.  In this approach the
``Lagrangian-history direct-interaction" approximation is just the first
step\cite{K965}. There is the price to pay for the elimination of the
sweeping in this theory:  The ``Lagrangian-history" perturbation expansion
has NO Feynman-type diagrammatic representation of $n$-th order terms in
the expansion.  Therefore it is very difficult to analyze high-order terms
in this approach and we do not know any example of its use in a theory of
hydrodynamic turbulence besides direct interaction approximations.

A different way to overcome the problem of infrared divergences due to the
sweeping is by the use of the Belinicher-L'vov resummation\cite{BL87} of
the Wyld diagrammatic expansion. This resummation corresponds to a change
of variables in the effective action of the Martin-Siggia-Rose
approach\cite{MSR3}  from the Eulerian velocity to the
quasi-Lagrangian (qL) velocity. The qL description of turbulence was
suggested by L'vov in 1980 (see\cite{BL87}).  The relation between the qL
velocity {\bf u} and the Eulerian velocity {\bf v} is
\begin{equation}
{\bf v}(t,{\bf r})={\bf u}(t,{\bf r}-\bbox{\varrho}(t))
\label{I1}
\end{equation}
where the time evolution of the vector $\bbox{\varrho}(t)$ is determined
by the equation
\begin{equation} d\bbox{\varrho}/dt$ $={\bf u}(t,{\bf r}_0)
\label{I2}
\end{equation}
where ${\bf r}_0$ is a marked point which may be arbitrarily chosen.  Let
us stress again that there are no approximations in this step; the above
relation is simply a change of variables in the   Martin-Siggia-Rose
effective action. A systematic analysis of resulting resummed series was
done in paper\cite{BL87}. In contrast to Kraichnan's Lagrangian-history
approach the resulting series do have a Feynman-type diagrammatic
representation. In contrast to the initial Wyld diagrammatic
technique\cite{W961} (and the Martin-Siggia-Rose one\cite{MSR3}) resummed
diagrammatic series do possess Galilean invariance order by order before
and after Dyson's line renormalization.  There is a price to pay for these
advantages: the transformation (\ref{I1},\ref{I2}) breaks the space
homogeneity of the problem. Therefore the Green's and correlation
functions in the qL representation depend on the space coordinates ${\bf
r}$ and ${\bf r}^\prime$ separately in contrast to standard approaches in
which these functions depend only on their differences.  However it was
possible to overcome the corresponding technicalities\cite{BL87}  and to
prove that there are no infrared nor ultraviolet divergences in any term
in any order of perturbation series.  As a result we have a powerful
technique which allows one to reach, as we believe,  crucial progress in
understanding the scaling of developed hydrodynamic turbulence.

The important step on that way was made in\cite{BL87}, (see
also\cite{L991}) where it was shown that the structure functions of
velocity differences with KO41 scaling are an order by order solution of
the corresponding diagrammatic equations.  In our 1993 paper\cite{LL93} we
show that the KO-41 scaling of velocity differences is the unique solution
(in some region of scaling exponents) under the very plausible assumption
that the time-dependent correlation functions of qL velocities are scale
invariant.  This is a non-perturbative result which follows from our
``frequency sum  rule for the dressed vertex"\cite{LL93}.  The later is an
exact consequence from the causality principle for the three-particle
Green's function.

If we do believe  that the Belinicher-L'vov resummation\cite{BL87,L991}
gives an adequate formalism to describe developed hydrodynamic turbulence
we should next discover in these terms the mechanism for anomalous
(non-Kolmogorov) scaling of the energy dissipation field. We have
introduced such a mechanism in brief \cite{LL94}.  In the direct qL
diagrammatic expansion for the correlation function of the energy
dissipation $K_{\varepsilon \varepsilon }(R)$ (\ref{J04}) we found an
infinite subset of logarithmically diverging diagrams. Their resummation
leads to a renormalization of the normal Kolmogorov-41 scaling behavior of
$K_{\varepsilon \varepsilon }(R)$. Let us stress that this is again a
non-perturbative result obtained with a diagrammatic technique.  There is
no closure procedure which leads to anomalous scaling; and only
resummation of an infinite series of relevant terms of the diagrammatic
perturbation expansion may produce this effect \cite{LL94}.  In ``physical
language" the mechanism responsible for the anomalous scaling was called
the {\it telescopic multi-step eddy interaction}.  The essence of this
mechanism is the existence of very large number $(R/\eta)^{\Delta_j}\gg 1$
of channels of interaction of large eddies of scale $R$ in the inertial
interval with eddies of viscous scale $\eta$ via a set of eddies of all
intermediate scales between $R$ and $\eta$.  Note that the same mechanism
works also for any local field constructed from the velocity gradients,
e.g., $\omega^2$ where $\bbox{\omega} = \bbox{\nabla}\times {\bf v}$ is
the {\it vorticity vector}. Our findings mean that KO-41 scaling describes
only a part of a full set of correlation functions characterizing
developed turbulence.

The next challenge for the resummed description of turbulence is to
understand the observable deviations of the scaling of structure functions
of velocity differences from the KO-41 prediction. Although a consistent
theory of this phenomenon belongs to the future a possible way to describe
these deviations as intermediate asymptotic behavior structure functions
at very large but finite Reynolds numbers is suggested by L'vov and
Procaccia in their ``subcritical scenario"\cite{LP94}.   If we accept this
scenario it would mean that all available experimental observations of the
scaling behavior of developed turbulence may be understood in the
framework of the resummed diagrammatic approach\cite{BL87}.

In the present paper  a consistent  theory of the anomalous scaling of the
energy dissipation field (and some other hydrodynamic fields constructed
from  velocity derivatives) is developed in the limit Re$\to\infty$. The
theory begin with the  Martin-Siggia-Rose approach\cite{MSR3}) to the
Navier Stokes equation and based on the Belinicher-L'vov resummation of
corresponding diagrammatic series.  It gives an analytical description of
the telescopic multi-step eddy interaction suggested in our
Letter\cite{LL94}.  We show that the correlation functions of the
hydrodynamic fields constructed from velocity gradients possess a
complicated behavior characterized by an infinite set of scaling
dimensions which do not reduce to the Kolmogorov ones.  Unfortunately
these exponents cannot be calculated explicitly since they describe the
scaling behavior of a complicated integral equation the kernel of which is
not found analytically.  Nevertheless a set of relations the between
scaling behaviors of correlation functions of different hydrodynamic
fields can be established.  We found a number of selection rules related
to the symmetries, including also the hypothetical conformal symmetry of
developed turbulence, and formulated some restrictions related to the
incompressibility condition of hydrodynamic motion. Our scheme leads also
to an anomalous asymptotic behavior of correlation functions of velocity
differences which are described by the same scaling exponents.

A tool which is useful for us is the set of so-called {\it fusion rules}
for fluctuating fields introduced by Polyakov \cite{P969} in the context
of second order phase transition theory.  These rules enable one to
determine the asymptotic behavior of a correlation function in the case
where a number of points (in the real {\bf r}-space) are separated from
other points by a large distance. Kadanoff \cite{K969} and Wilson
\cite{W969} formulated these rules in the form of the so-called {\it
operator algebra}. This algebra was extensively used in the conformal
theory of $2d$ second order phase transitions \cite{BPZ4}. Since the
algebra is not related to any particular properties of the system one may
expect that it can be formulated for any system described by a set of
correlation functions with a scaling behavior.  In this work we are going
to propose an operator algebra for $3d$ turbulence, which is characterized
by a set of correlation functions of fluctuating turbulent fields.

The structure of the paper is as follows. Section \ref{sec:kolob} contains
a brief overview of the basic physical concepts of hydrodynamic turbulence
associated with the idea of scaling.  In Section \ref{sec:teles} we
develop the qualitative picture  of the ``telescopic multi-step
mechanism'' of eddy interaction suggested in our 1994 Letter\cite{LL94}.
This allows one to recognize physical reasons for the anomalous scaling
behavior of the energy dissipation field and other hydrodynamic fields.
In Section \ref{sec:diagr} we develop the analytical theory beginning with
the Navier-Stokes equation in the quasi-Lagrangian representation and
based on a line-renormalized diagrammatic approach.  We extract infinite
sub-sequences of diagrams with ultraviolet logarithmic divergences which
lead after summation to anomalous scaling behavior of the correlation
functions of the hydrodynamic fields with different scaling exponents. In
Section \ref{sec:fusrl} we formulate rules, including the fusion rules,
which allow us to predict scaling behavior of correlation functions of
hydrodynamic fields constructed from the velocity gradients and of the
velocity differences. The results obtained and a discussion concerning the
region of applicability of our approach are presented in the Conclusion.

\section{Scaling in Hydrodynamic Turbulence}
\label{sec:kolob}
In this section we recall the main ideas associated with the scaling
behavior of different correlation functions of the hydrodynamic fields
such as velocity differences and fields constructed from the velocity
gradients. We introduce the notion of normal scaling behavior determining
by the Kolmogorov's dimension estimates and also anomalous exponents which
describe deviations from this normal behavior.
\subsection{Kolmogorov 1941 Scaling for Turbulent Velocity}
Let us begin by accepting The Kolmogorov's 1941 picture of turbulence
(KO-41) and consider within this framework the statistical properties of
the turbulent velocity field $v_\alpha(t,{\bf r})$.  The principal
quantity characterizing the turbulence in KO-41 is the average value
$\bar\varepsilon$ of the energy dissipation rate
\begin{equation}
\varepsilon(t,{\bf r}) =2 \nu s^2(t,{\bf r})\,,
\label{dis}
\end{equation}
where $s^2$ is the second power $s^2=s_{\alpha\beta}s_{\beta\alpha}$ of
the {\it strain tensor}: $s_{\alpha \beta} = \big( \nabla_\alpha
v_\beta+\nabla_\beta v_\alpha\big)/2$.  Note that $\varepsilon$ is the
energy dissipation rate per unit mass; to find the energy dissipation rate
per unit volume $\varepsilon$ should be multiplied by the mass density
$\rho$.

We should be careful in treating correlation functions containing the
velocity field ${\bf v}(t,{\bf r})$ itself. First, this quantity is not
Galilean invariant and depends on the choice of the reference system.
Second, the main contribution to correlation functions of velocities (like
$\langle{\bf v}(t,{\bf r})\cdot {\bf v}(t,{\bf r}^{\prime})\rangle$) comes
from the largest eddies. This means that expressions for such correlation
functions contain large homogeneous contributions which cannot be obtained
in terms of the fields defined in the inertial interval. As a result such
correlation functions are not universal: they may depend on boundary
conditions, statistics of pumping etc. In the analytical approach the
dominating contribution to correlation functions associated with the
energy-containing subrange is reflected in infrared divergences of
corresponding integrals. It is well known that to avoid this difficulty we
should consider such objects as velocity differences which are Galilean
invariant; their correlation functions  do not contain infrared terms
related to the largest eddies.

Let us introduce the traditional designation for the average values of the
powers of the velocity difference:
\begin{equation}
D_n(R)\equiv\langle \mid  {\bf v}(t,{\bf r}) -
{\bf v}(t, {\bf r}+{\bf R}) \mid^n \rangle
\label{J07} \,.
\end{equation}
The quantities $D_n$ are usually called the {\it structure functions of
the velocity differences}.  We will be interested in the behavior of these
structure functions in the inertial subrange $L\gg R\gg\eta$.  In
Kolmogorov's picture one can evaluate $D_n(R)$ as $\bar\varepsilon^a\,R^b$
with the exponents $a$ and $b$ chosen in such a way as to have the same
dimensionality (remind that $\bar\varepsilon$ is the energy dissipation
rate per unit mass). This condition gives $a=b=n/3$, which leads to the
famous KO-41 result $D_n(R)=C_n (\bar\varepsilon R)^{n/3}$ with some
dimensionless constants $C_n$. Physically these dimensional estimates mean
that the principal contribution to the functions $D_n(R)$ is associated
with $R$-eddies having characteristic velocity $V_{_R}$ of the order of
\begin{equation}
V_{_R}\sim (\bar\varepsilon R)^{1/3} \label{veloc} \,.
\end{equation}
The characteristic turnover time of $R$-eddies in KO-41 can be estimated
as $R/V_{_R}$. Since in KO-41 one can construct only one quantity with the
dimensionality of time, the lifetime of $R$-eddies $\tau_{_R}$ (at $R$ in
the inertial subrange) should be of the order of their turnover time.

In treating correlation functions of the velocity gradients one has no
problems with large-scale terms since these correlation functions are not
sensitive to motions on the pumping scale $L$. KO-41 dimensional reasoning
allows one to find the scaling behavior of the two-point correlation
function of the strain tensor $s_{\alpha\beta}$ and the vorticity vector
$\bbox{\omega}=\nabla\times{\bf v}$. For this we can use (\ref{veloc})
which gives for eddies of the scale $R$ that  $s,\omega \sim\bar
\varepsilon^{1/3} R^{-2/3}$. Moreover, one can establish the tensor
structure of the pair simultaneous correlation functions of $s,\omega$
starting from
\begin{eqnarray}
D_{\alpha\beta}({\bf R})&\equiv&
\langle(v_\alpha({\bf R})-v_\alpha(0))
(v_\beta({\bf R})-v_\beta(0))\rangle
\nonumber \\
 &=& \frac{4}{11}C_2 {\bar\varepsilon R}^{2/3}
\big(\delta_{\alpha\beta}- R_{\alpha}R_{\beta}/4R^2\big)  \ .
\label{str}
\end{eqnarray}
The tensorial structure here is determined by the incompressibility
condition $\nabla\cdot{\bf v}=0$.\  $C_2$ is a dimensionless constant.
The pair correlation functions of the velocity differences can now be
found  as
\begin{equation}
\langle\nabla_\mu v_\alpha({\bf R})
\nabla_\nu v_\beta(0)\rangle = \frac{1}{2} \nabla_\mu\nabla_\nu
D_{\alpha\beta}({\bf R}) \ .
\label{PC}
\end{equation}
\subsection{Normal and Anomalous Scaling in Turbulence}
In the previous Section the ``na\"{\i}ve'' KO-41 scaling exponent for some
correlation functions have been introduced. We refer to these exponents as
{\it normal}.  As we noted strong fluctuations could change these
``na\"{\i}ve'' exponents. We refer to the deviations from KO-41 values as
{\it anomalous} exponents. Accordingly we introduce in this Section the
formal designations for the anomalous exponents characterizing different
correlation functions.

Intermittency is associated with strong fluctuations of the energy
dissipation rate $\varepsilon$. To characterize these fluctuations we
should treat correlation functions of $\varepsilon$.  Introduce the
designation for the pair correlation function
\begin{equation}
K_{\varepsilon \varepsilon }(R)\equiv
\langle \langle\varepsilon (t, {\bf r})
\varepsilon(t, {\bf r}+{\bf R)}\rangle\rangle \ .
\label{J04}
\end{equation}
Here and henceforth the double angular brackets mean the irreducible part
of the correlation function of two fields:
$$
\langle\langle\psi({\bf r}_1)\varphi({\bf r}_2)\rangle\rangle\equiv
\langle \psi({\bf r}_1)\varphi ({\bf r}_2)\rangle
-\langle \psi \rangle\langle \varphi \rangle\ .
$$
It is natural to assume\cite{MY75} that the correlation function
$K_{\varepsilon\varepsilon}(R)$ is scale invariant in the inertial
subrange:
\begin{equation}
K_{\varepsilon\varepsilon}(R) \propto R^{-\mu}
\label{J06}
\end{equation}
with some phenomenological exponent $\mu$. This quantity is usually called
the {\it intermittency exponent}.

There are two known ways to evaluate the exponent $\mu$ within dimensional
reasoning.  The simplest one is as follows.  Clearly the dimensionality of
$K_{\varepsilon\varepsilon}(R)$ coincides with the dimensionality of
$\varepsilon^2$. Therefore an attempt to express $K_{\varepsilon
\varepsilon}$  via powers of $\varepsilon$ and $R$ gives $\mu=0$
\begin{equation}
K_{\varepsilon\varepsilon}(R) \sim  \bar \varepsilon^2 R^0  \ .
\label{E1}
\end{equation}
In contradiction with the folklore (see, for example\cite{MY75}) this
estimate is not related to KO-41.  Indeed, according to (\ref{dis}) the
energy dissipation field $\varepsilon(t,{\bf r}) $ is proportional to the
viscosity $\nu$ which is not a value characterizing the inertial subrange
and it is not reasonable to rewrite the parameter $\nu$ characterizing the
medium in terms of $\bar\varepsilon$ and $r$.  To be consistent with KO-41
we should treat the correlation functions independently of $\nu$, in terms
of $s^2$.  Then according to KO-41 dimensional reasoning
\begin{equation}
\langle \langle s^2 ({\bf r})
s^2({\bf r}+{\bf R)}\rangle\rangle
\sim\bar\varepsilon^{4/3}R^{-8/3}
\label{E3} \,.
\end{equation}
Since $K_{\varepsilon \varepsilon}= 4 \nu^2 s^2s^2$ this expression means
that $\mu=8/3$. The explicit expression for $K_{\varepsilon \varepsilon}
(R)$ with $\mu=8/3$ was found by Golithin \cite{G962} under the assumption
of Gaussianity of the velocity fluctuations. As we show one can obtain it
within the ``spirit" of KO-41.  We see now that the normal KO-41 exponent
leads to an extremely fast decay of $\varepsilon \varepsilon$
correlations.

During last three decades a lot of effort has been expended in attempting
to measure the exponent $\mu$ (see e.g.
\cite{GZ63,PS65,GSC0,TW72,FK75,McC6,P976,VA80,AGH4}). There are many
experimental difficulties in this problem, therefore different authors
give for $\mu$ values ranging from $0.1$ to $0.5$. Thus it is clear that
the normal KO-41 exponent $\mu=8/3$ does not correspond to experiment.
This is not amazing at all, because from a consistent theoretical
viewpoint there is no reason to accept both Eq.  (\ref{E3}) giving
$\mu=8/3$ and  Eq. (\ref{E1}) giving $\mu=0$.  Note that the multifractal
pictures of turbulence with some assumption about the structure of the
flow\cite{CLT0,She4} give for the exponent $\mu$ an acceptable value near
$1/3$.  We conclude that the correlation functions (\ref{J04}) undoubtedly
possess an anomalous behavior.

The physical reason for deviations of the correlator $K_{\varepsilon
\varepsilon}(R)$ from (\ref{E3}) is clear, see for
example\cite{F991,BAFG}. Namely, the $s^2$-, and the $\omega^2$-fields as
well as $s^{2n}$-fields and many others, are so-called {\it viscous
subrange fields} in the sense that mean values of these fields are mainly
determined by eddies of characteristic scale $\eta$.  It means that the
$\eta$-eddies may play an essential role in forming the behavior of the
correlation functions of these fields.  Therefore the correlation
functions of these fields have to be not only a function of the inertial
subrange values as in the normal KO-41 estimate (\ref{E3}), but also a
function of the viscous scale $\eta$. As a result powers of the
dimensionless parameter $R/\eta$ can enter as a factor in all correlation
functions of $s^{2n}$ and $s^{2n}$-fields. There is no dimensional
reasoning which can determine to what power this parameter appears. So,
the only result which follows from a consistent dimensional reasoning is
\FL
\begin{equation}
\!\langle\langle s^{2n}({\bf R}) s^{2m}(0) \rangle\rangle=
C_{nm} \Big({R\over\eta}\Big)^{\Delta_{nm}} \!\!
\left(\frac{\bar\varepsilon}{R^2}\right)^{2(m+n)/3}
\label{est1}
\end{equation}
with some unknown exponents $\Delta_{nm}$.  Note that both particular
cases (\ref{E1}) and (\ref{E3}) are described by (\ref{est1}) for some
values of $\Delta_{11}$.  We will consider the relations (\ref{est1}) as a
definition of the anomalous scaling exponents $\Delta_{nm}$ which
determine the scaling behavior of the prefactor of the na\"{\i}ve KO-41
expression for the correlation function. Exactly the same problem with
unknown dimensionless factors arises for many-point correlation functions
of other viscous subrange fields.  An important question concerns also the
dimensionless factors $\eta/R$ and $R/L$ in the structure functions of
velocity differences (\ref{J07}).  The appearance of such factors is not
forbidden from the general point of view, if to include these factors
besides KO-41 estimates we find for (\ref{J07}):
\begin{equation}
D_n(R)\propto R^{1/3}\left({\eta\over R}\right)^
{\Delta_n}\left({ R\over L}\right)^{\tilde\Delta_n} \propto R^{\zeta_n}\,,
\label{J08}
\end{equation}
where the scaling exponents $\zeta_n$ differ from the KO-41 prediction
$n/3$.  In his lognormal model of intermittency \cite{K962} Kolmogorov
argued that
\begin{equation}
\zeta_n=n/3-\mu n (n-3)/18 \ .
\label{LNM}
\end{equation}
In the Novikov-Steward \cite{NS64} model and in the $\beta$-model by
Frisch, Sulem, and Nelkin\cite{FSN8}
\begin{equation} \zeta_n=n/3- \mu (n-3)/3 \ .
\label{BETA}
\end{equation}
Here $\mu$ is the same scaling exponent as in $K_{\varepsilon
\varepsilon}(R) \propto R^{-\mu}$. Note that in these models the deviation
of $\zeta_n$ from $n/3$ is related to the decay of $\varepsilon
\varepsilon$ correlations. In fact, there is no solid basis for this
statement \cite{MS91,F993}.

Actually we do not see serious theoretical reasons for the deviations of
$\zeta_n$ from their KO-41 values $n/3$ in the limit Re$\to \infty$.
Indeed,  on one hand there is the theorem proved by Belinicher and L'vov
\cite{BL87,L991} who demonstrated that the KO-41 structure functions of
velocity differences order by order satisfy the corresponding diagrammatic
equations.  Moreover, in 1993 \cite{LL93} we showed that KO-41 is the
unique solution (in some region of scaling exponents) under the assumption
that the time-dependent correlation functions of qL velocities are scale
invariant.  On the other hand we do not know any mechanism for the
renormalization of the scaling in the absence of ultra-violet and infrared
divergences  of diagrams in nonlinear problems with strong interaction.
Therefore in the present paper we have accepted the KO-41 scaling of
velocity differences ($\zeta_n=n/3$) and find some remarkable consequences
of this fact concerning the anomalous exponent of different correlation
functions of the velocity gradients as introduced in (\ref{est1}).
Certainly there are deviations from $\zeta_n=n/3$ which are observed
experimentally (see, e.g., \cite{BCTB,SSJ3} and references therein) and in
numerically\cite{BO94}.  However we believe that these deviations are
related to the finite value of Re in experiments  which leads to
restricted values of the inertial subrange $L/\eta < 10^4$ for the largest
available Reynolds numbers ${\rm Re}< 10^8-10^9$ (see also Conclusion).  A
possible ``subcritical" mechanism for such intermediate non-Kolmogorov
behavior was suggested recently by L'vov and Procaccia\cite{LP94}.
\section{Telescopic Multi-Step Eddy Interaction.}
\label{sec:teles}
Our aim in this Section is to evaluate the correlation functions of $s^2$
and $\omega^2$ for a separation distance $R$ within the inertial subrange
$L\gg R \gg \eta$. We show that the leading contributions to the
correlation functions $K_{s2,s2}(R)$, $K_{s2,\omega 2}(R)$, and $K_{\omega
2,\omega 2}(R)$ have anomalous scaling behavior characterized by the same
scaling exponent $\Delta_1$. We argue that at the same time the
correlation function of a traceless tensor (say $\omega_\alpha
\omega_\beta -\case{1}{3}\delta _{\alpha\beta}\omega^2$) has to have
another independent anomalous scaling exponent.  The  {\it telescopic
multi-step eddy interaction} mechanism for the anomalous exponent of the
energy dissipation field was suggested in our Letter \cite{LL94} through
the diagrammatic approach.  In order to elucidate the physical basis of
the involved diagrammatic calculation we first describe  our findings in
this Section using the popular handwaving language of cascades, eddies and
their interactions before turning to the cumbersome technicalities of the
analytical theory of turbulence.

In our approach $K_{s2,s2}$ and other correlation functions are
represented as a sum over the contributions from turbulent fluctuations in
the inertial range. The fluctuations $s^2({\bf r})$ and $\omega^2({\bf
r})$ are due to fluctuations of the velocity gradient at point ${\bf r}$
which in turn may be regarded as the result of a superposition of shear
rates stemming from velocity differences on various length scales $r$,
which according to KO-41 scale like $r^{-2/3}$. Therefore the main
contribution to the fields of interest themselves is expected to come from
the smallest scales near $\eta$ while the main contribution to the
correlation over the distance $R$ is na\"{\i}vely expected from the
smallest scales which bridge the gap between ${\bf r}$ and ${\bf r}+{\bf
R}$, namely from the scales $R$.  This means that scales larger and
smaller than $R$ can be neglected in our considerations.  We call the
above reasoning {\it na\"{\i}ve}\/ because it tacitly assumes that an
indirect action through scales $<R$ is much weaker.

In this Section first we consider the simplest contribution to the
correlation function of our fields coming from eddies of one scale $R$.
Than we show that this is not the whole story  and that the combined
effect of indirect interactions of scale $R$ with scale $\eta$ is also
important.  We will examine the contribution from two and three groups of
eddies of different scales. Finally we consider the total contribution of
eddies of all scales from $\eta$ to $R$ and show that the indirect
interactions of scale $R$ with scale $\eta$ via  the set of all
intermediate scales gives rise to a correction of the scaling exponents of
our fields from Kolmogorov's values downward.
\subsection{Direct Contribution of One-Scale Eddies}
In order to formally determine the notion of $x$-eddies let us partition
$k$-space into shells separated by wavenumbers $k_n=2\pi / x_n$. By
$n$-eddies (or eddies of scale $x_n$) we mean the turbulent velocity field
${\bf v}_n(t,{\bf r})$ which has only  part of the Fourier harmonics of
the ``real" turbulent velocity field ${\bf v}(t,{\bf k})$ with $k$ between
$k_{n-1}$ and $k_n$.

Consider now the contribution to $K_{s2,s2}(R)$ due to eddies of some
characteristic scale $x_n$.  It is clear that for a separation distance
$R$ which is about the eddy scale $x_n$, the  correlation function
$K_{s2,s2}(R)$ has to be of order of the KO-41 estimate (\ref{E3}).
Indeed, in this case $R$ and $\bar\varepsilon$  are the only parameters in
the problem considered.

In the case of $x_n \gg R$ (but still $x_n \ll L$) the correlator $K_{s2,
s2}(R)$  is independent of $R$ and is approximated by $ K_{s2,s2}^{^{\rm
k41}}(x_n)\propto x_n^{-8/3}$, see (\ref{E3}).  Thus this estimate of
$K_{s2,s2}(R)$ is smaller than (\ref{E3}).

For $x_n \ll R$ the correlator $K_{s2,s2}(R)$ has to be exponentially
small with respect to (\ref{E3}). Therefore for a one-scale contribution
the largest estimate of $K_{\varepsilon\varepsilon}(R)$ is given for $x_n
\simeq R$ and corresponds to the KO-41 value (\ref{E3}).
\subsection{Contribution of One-Step Eddy Interactions}
Consider now the contributions of two group of eddies (or scales $x_n $
and $x_m$) homogeneously distributed in space. First of all there are the
contributions of each group of eddies separately.  Second there is a
cross-contribution arising from the interaction of these two groups of
eddies. By discussions similar to those presented in the previous
Subsection one may see that the largest cross-contribution to
$K_{s2,s2}(R)$ in this case appears at $x_m \simeq R$, and an arbitrary
value of $x_n $ (or vice versa).  Therefore we choose from the beginning
$m=N$ with an arbitrary value of $n$.

Let us consider correlations between fluctuations on different scales with
the help of a conditional probability density. The larger eddies modify
the statistics of the smaller ones locally in $r$-space by their
rate-of-strain field. Thus the probability density to find on scales $\leq
n$ near $\bf r$ the velocity gradient $\nabla{\bf v}$ is
$
P_n\left(\nabla{\bf v}({\bf r}+{\bf r'})| \nabla({\bf v}_{n+1}({\bf r})+
{\bf v}_{n+2}({\bf r})+\dots)\right)
$,
where the condition is set by all the larger scales and $|{\bf r'}|\ll
x_{n+1}$. We assume that fluctuations around the local mean value at
points farther apart than $|{\bf r'}|\ll x_{n+1}$ are not correlated.
Further $\lfloor\dots\rfloor_n$ will denote such conditional averages.
Obviously, $\big\lfloor\lfloor\dots\rfloor_m \big\rfloor_n =
\lfloor\dots\rfloor_l$ holds with $l = \max(m,n)$.  KO-41 scaling implies
that the velocity gradients of larger scales are comparatively small.
Hence, we will make use of the expansion of $P_n$ for small large-scale
gradients, $P_n(\nabla{\bf v}| \nabla{\bf v}_{>n}) = P_n^{(0)}(\nabla{\bf
v}) +P_n^{(1)}(\nabla{\bf v})\cdot\nabla{\bf v}_{>n} +P_n^{(2)}(\nabla{\bf
v})\cdot\nabla{\bf v}_{>n}\nabla{\bf v}_{>n}$.

We illustrate in this subsection the consequences of such correlations
considering the hypothetical case where only fluctuations in shell $N$ and
in shell $n$, $1\ll n\ll N$ are excited. Then
\begin{eqnarray}
K_{s2,s2}(|{\bf r}_1-{\bf r}_2|)&=&\left\langle
\left\langle  s^2(t,{\bf r}_1)
s^2(t,{\bf r}_2)\right\rangle \right\rangle
\nonumber \\
\equiv
\left\langle   \widehat{s^2}({\bf r}_1,t)  \widehat{s^2}{\bf r}_2,t)
\right\rangle
&=&\Big \lfloor\big \lfloor
 { s^2}(t,{\bf r}_1)  { s^2}(t,{\bf r}_2)
\big \rfloor_n  \Big \rfloor_{_N}
\label{AV} \\
&=& \Big \lfloor \big \lfloor
\widehat{  s^2}(t,{\bf r}_1) \big\rfloor_n
\big\lfloor \widehat{  s^2}(t,{\bf r}_2)
\big\rfloor_n  \Big\rfloor_{_N}  \ .
\nonumber
\end{eqnarray}
where we denote $s^2(t,{\bf r})-\langle s^2 \rangle$ as $\widehat{
s^2}(t,{\bf r})$. The last step in (\ref{AV}) is justified because the
$s^2$ field is mostly sensitive to the fast, small scale motions; each
$x$-ensemble average can be done in the presence of some realization of
the $R$-eddies, and only when we compute the correlation function
(\ref{AV}) we need to average in the $R$-ensemble.  In other words one may
split the averaging in the above equation because the correlation length
under the conditional $n$-average is much shorter than $R$.

Consider now the average over small $n$-eddies
\begin{equation}
 \big\lfloor\widehat{ s^2}(t,{\bf r}) \big\rfloor_n
= \big\lfloor
s^2(t,{\bf r}) - \left\langle s^2  \right\rangle
\big \rfloor_n \ .
\label{AV1}
\end{equation}
Using the expansion of the conditional probability we obtain for this
object an expression of the form
\begin{eqnarray}
\big \lfloor \widehat{ s^2} (t,{\bf r}) \big \rfloor_n &\simeq&
s^2_{_N}(t,{\bf r}) + A_{\alpha\beta,n}
\nabla_\alpha v_{\beta,_N} (t,{\bf r})
\nonumber \\
&&+ B_{\alpha\beta\gamma\delta,n}\,
\nabla_\alpha v_{\beta,_N} (t,{\bf r})
\nabla_\gamma  v_{\delta,_N}(t,{\bf r})
\label{B3}\\
&&+ {\rm higher~~order~~terms}     \ .
\nonumber
\end{eqnarray}
The first term on the RHS of this equation is the direct contribution of
the $N$-shell to $\lfloor\tilde{\varepsilon}({\bf r})\rfloor_n$.  The
largest contribution to $\lfloor s^2 ({\bf r})\rfloor_n$ comes from the
$n$-shell itself, and is  $ \lfloor s^2_n( {\bf r})\rfloor_n$.  However
this contribution is independent of time and space coordinates and is
canceled in the subtraction of $\bar \varepsilon$. The last two terms
derive immediately from the expansion of $P_n$.  The term of first order
in $\nabla{\bf v}$ is not important because all contributions originating
from this term will vanish finally under the average over fluctuations on
intermediate scales.  The expansion coefficients $B$ are the corresponding
derivatives of the function $\lfloor s^2(t,{\bf r}) \rfloor_x$ taken at
zero value of ${\bf v}_{_N}(t,{\bf r})$.  Therefore they reflect the
``intrinsic" properties of $n$-scale turbulence, without interaction with
$R$-eddies (that is the reason why we display in (\ref{B3}) the dependence
of $A$ and $B$ on the scale $x_n$). Note that the turbulence of $x$-eddies
is isotropic and because of this the matrices $ A$ and $ B$ have to have
some particular form which allows one to represent the above expansion in
a more elegant form:
\begin{eqnarray}
\lfloor \widehat{ s^2} (t,{\bf r}) \rfloor_n
&\simeq &
\widehat{ s^2_{_N}}(t,{\bf r})
\nonumber \\
&&+   B_{11,n} \widehat{ s^2_{_N}}(t,{\bf r}) +  B_{12,n}
\widehat{ \omega^2_{_N}}(t,{\bf r}) +...
\label{B6}
\end{eqnarray}
where $s^2_{_N}(t,{\bf r})$ and $\omega^2_{_N}(t,{\bf r})$  are the
contributions to the square of the strain tensor and the vorticity coming
from the $N$-shell and ``~$\widehat{{~~~}}$~" denotes the irreducible part
in accordance with (\ref{AV1}). It is very easy to understand why one has
only two terms ($\propto s^2$ and $\propto \omega^2$) in the RHS of
(\ref{B6}): the LHS of this equation is scalar (under all rotations)
therefore the RHS also has to have a scalar form, and there are two
scalars ($s^2$ and $\omega^2$) which one can build from $(\nabla_\alpha
v_\beta)(\nabla_\gamma v_\delta)$. Clearly, the expansion for another
scalar $\lfloor \omega^2 (t,{\bf r}) \rfloor_x$ has to be similar to
(\ref{B6}):
\begin{eqnarray}
\big \lfloor \widehat{
\omega^2 }(t,{\bf r}) \big \rfloor_n
&=& \widehat{ \omega^2_{_N}} (t,{\bf r})
\nonumber \\
&& +
  B_{21,n}\widehat{  s^2_{_N}}(t,{\bf r}) +  B_{22,n}
\widehat{ \omega^2_{_N}}(t,{\bf r}) +...
\label{B7}\ .
\end{eqnarray}
One may diagonalize the matrix $B_{ij}$ and find linear combinations of
fields $s^2$ and $\omega^2$
\begin{eqnarray}
\Psi_1(t,{\bf r})&=&U_{11}\widehat{ s^2}
+U_{12}\widehat{ \omega^2}\,,
\nonumber\\
\Psi_2(t,{\bf r})&=&U_{21}\widehat{ s^2}+U_{22}\widehat{ \omega^2}\ .
\label{B8}
\end{eqnarray}
for which the expansions (\ref{B6},\ref{B7}) (up to second order
terms in $\nabla v$) take the simplest form:
\begin{eqnarray}
&& \big \lfloor \Psi_j(t,{\bf r}) \big
\rfloor_n
\label{B9}\\
&=&  \Psi_{j,N}   (t,{\bf r}) \big[ 1 +
B_{j,n} \big] \,,\quad j=1\,,2\ .
\nonumber
\end{eqnarray}
The coefficients $B_{j,n}$ can  be estimated applying physical reasoning
(or, equvalently, the Navier Stokes equations). The dimensionless
parameter describing the relative change of velocity ${\bf v}_n( {\bf r})$
with varying $\nabla{\bf v}_{_N}$ is $\tau_n\nabla{\bf v}_{_N}({\bf r})$,
where $\tau_n$ is the life time of $n$-eddies. Therefore, $ B_{j,n} \sim
\tau_n^2 \lfloor |\nabla v_n|^2 \rfloor_n \sim x_n^{0}$. Here we have used
essentially the KO-41 result that the life time $\tau_n$ is approximately
the turnover time of the $n$-eddies. This allows one to conclude that
$B_{j,n}$ is independent of the scale $x_n$. This statement is of crucial
importance for the understanding of the origin of the anomalous scaling:
all scales in the interval from $R$ to $\eta$ contribute  equally to the
correlation functions under consideration.  Therefore we have to take into
account not only contributions from two group of eddies, as we did up to
now, but contributions of all the group of eddies of scales from the above
interval.
\subsection{Contributions of Two- and Many-Step Eddy Interactions}
Now let us consider the contribution of  three groups of eddies (with
scales $R$, $x_n$, and $x_m$). This will directly lead to anomalous
scaling. For this contribution one has:
\begin{eqnarray}
&&\lfloor \Psi_j(t,{\bf r}) \rfloor_n
\label{J09} \\
&=& \Psi_{j,N}(t,{\bf r})
\big(1+B_{j,n}+B_{j,m}+B_{j,n}\,B_{j,m} \big) \ .
\nonumber
\end{eqnarray}
The first three terms on the RHS of this equation describe the direct
contribution of the $R$-shell and the contributions from the direct
interaction of the $R$-shell with the $n$- and $m$-shell,
respectively.  The last term ($\propto B_{j,n}\,B_{j,m} $) is due to the
indirect effect of the largest scale, the $R$-shell, on the smallest
scale, the $n$-shell say, via the intermediate $m$-shell. To obtain
this term one has to repeat twice the above expansion. The RHS of (\ref{J09})
is proportional to $(1+B_{j,n})(1+B_{j,m})$. Thus
it is plausible that one obtains in the case of $N$ shells the result
\begin{equation}
\lfloor \Psi_j(t,{\bf r}) \rfloor_{_N}  \simeq
\Psi_{j,_{N}} (t,{\bf r})
\prod_{n=1}^N
\big(1+B_{j,n}\big)\ .
\label{J10}
\end{equation}
Note that the independence of the $B_{j,n}$ of $n$ as pointed out above
will only hold when the width $\Delta k_n$ of the shells scales in
the same way as the $k_n$ themselves. We choose $k_{n+1}/k_n=\Lambda$
so that neighboring shells may be considered as almost statistically
independent. Such $\Lambda>1$ does exist because of the locality of
energy transfer via the scales\cite{BL87}.  Then one can write
(\ref{J10}) in the form
\begin{eqnarray}
\lfloor \Psi_j(t,{\bf r}) \rfloor_{_N}  &\simeq&
\Psi_{j,_{N}} (t,{\bf r})
\big(1+B_j\big)^N
\nonumber  \\
&\simeq&
\Psi_{j,_{N}} (t,{\bf r}) \big(R/\eta\big)^{\Delta_j}
\label{J11}
\end{eqnarray}
where $N={\bf log}_\Lambda(R/\eta)$, $\Delta_j=\ln(1+B_j)/\ln (\Lambda)$
 and $j=1,\, 2$. Following the terminology of second order phase
transitions one can call the exponents $\tilde \Delta_1$ and $\tilde
\Delta_2$ {\it anomalous dimensions}  of the fields $\Psi_1$ and $\Psi_2$.
The physical meaning of $\tilde \Delta_j$ is clear from (\ref{J11}): the
factor $(R/\eta)^{\tilde \Delta_j}$ is the total number of effective
channels of multi-step eddy interactions of the $\Psi(R|t,{\bf r})$ field
(having scale $R$) with the smallest $\eta$-eddies.

Returning to the original fields $s^2$ and $\omega^2$ one has instead
of (\ref{J11}):
\begin{eqnarray}
&&\Big \lfloor\big \lfloor\dots \lfloor
\widehat{ s^2}(t,{\bf r}) \rfloor_y \big \rfloor_x \dots \Big
\rfloor_\eta
\nonumber \\
&\simeq& A_{11}\Psi_{1,_N}(t,{\bf r})
\Big(\frac{R}{\eta}\Big)^{\tilde \Delta_1}
+ A_{12}\Psi_{2,_N}2(t,{\bf r})
\Big(\frac{R}{\eta}\Big)^{\tilde \Delta_2}  \,,
\nonumber \\
&&\Big \lfloor\big \lfloor\dots \lfloor
\widehat{ \omega^2}(t,{\bf r})\rfloor_y \big \rfloor_x \dots \Big
\rfloor_\eta
\label{B15} \\
&\simeq& A_{21}\Psi_{1,_N}(t,{\bf r})
\Big(\frac{R}{\eta}\Big)^{\tilde \Delta_1}
+ A_{22}\Psi_{2,_N}(t,{\bf r})
\Big(\frac{R}{\eta}\Big)^{\tilde \Delta_2}
\nonumber
\end{eqnarray}
with some dimensionless coefficients $A_{ij}$.
\subsection{Additional Anomalous Fields and Correlation Functions}
Consider now correlation functions of the pseudovector field ${\bf
f}^*$$=s_{\alpha\beta} \omega_\beta$. Clearly, one can expand $\lfloor
{\bf f}^*(t,{\bf r}) \rfloor_x$ in a similar way to (\ref{B6}).
Previously we expanded a scalar in terms of scalars.  The pseudovector
field has to be expanded in the terms of pseudovector fields. Generally we
have to expand a field with  given transformation properties (under all
rotations and inversion) in terms of fields with the same transformation
properties. This requirement is a consequence of the isotropy of
turbulence and of the fact that the expansion coefficients reflect the
properties of the fine scale turbulence only.  So,
\begin{equation}
\lfloor {\bf f}^*(t,{\bf r}) \rfloor_x\simeq {\bf f}^*_{_N}(t,{\bf r})
[1+ b^* +\dots]\ .
\label{B17}
\end{equation}
Here $b^*$  is some new dimensionless expansion coefficient. Repetition
of all of above considerations allows one to conclude that
\begin{equation}
\Big \lfloor\big \lfloor\dots \lfloor
{\bf f}^*(t,{\bf r}) \rfloor_y \big \rfloor_x \dots \Big
\rfloor_\eta \simeq {\bf f}^*_{_N}(t,{\bf r}) \Big(\frac{R}{\eta}\Big)^{
\Delta^*}\,,
\label{B18}
\end{equation}
with new anomalous scaling exponents $\Delta^*\propto \ln (1+b^*)$. In the
same way one can find three new anomalous scaling exponents, associated
with three traceless tensor fields of the second order (in $\nabla {\bf
v}$),  one new exponent for a pseudotensor of the third order and so on.
For more detail, see Section IVB.

Now one can proceed to establish the scaling behavior of the two-point
correlation functions of the $s^2$-, $\omega^2$-, $f^*_\alpha$-, etc.
fields. For example, Eqs. (\ref{B15}) together with (\ref{AV}) allow one
to see that
\begin{eqnarray}
K_{s2,s2}(R)&\simeq& \frac{\bar \varepsilon^{4/3} }{R^{8/3}}
\Bigg[A_{11}^2 \Big(\frac{R}{\eta}\Big)^{2 \tilde \Delta_1}
\label{B16} \\
&+& 2A_{11} A_{12}\Big(
\frac{R}{\eta}\Big)^{\tilde \Delta_1+\tilde \Delta_2}
+A_{22}^2\Big(\frac{R}{\eta}\Big)^{2\tilde \Delta_2}\Bigg]\,,
\nonumber \\
K_{s2,\omega2}(R)&\simeq& \frac{\bar \varepsilon^{4/3} }{R^{8/3}}
\Bigg[A_{11}A_{21} \Big(\frac{R}{\eta}\Big)^{2 \tilde \Delta_1}
\label{B19} \\
+\big(A_{11} A_{22}&+&A_{12} A_{21}\big)\Big(
\frac{R}{\eta}\Big)^{\tilde \Delta_1+\tilde \Delta_2}
+A_{12}A_{22}\Big(\frac{R}{\eta}\Big)^{2\tilde \Delta_2}\Bigg]\,,
\nonumber
\end{eqnarray}
and a similar equation for $K_{\omega2,\omega2}$. The leading term in each
of these expressions scales like $R^{2\tilde \Delta_1- 8/3}$.  Expression
(\ref{B15}) also allows one to estimate in the same manner
cross-correlation functions of $s^2$ and $\omega^2$ with other
hydrodynamic fields, to estimate the ${\bf f}^*_1{\bf f}^*_2$-correlation
function
\begin{equation}
K_{{\bf f}*,{\bf f}*}=\left\langle\left\langle
{\bf f}^*({\bf r}_1){\bf f}^*({\bf r}_2)
\right\rangle \right\rangle
\sim \frac{\bar\varepsilon^{4/3}}{R^{8/3}}\Bigg(
\frac{R}{\eta}\Bigg)^{\tilde\Delta^*}+\dots\,,
\label{B20}
\end{equation}
etc. We will discuss  the problem of the evaluation of the correlation
function of various hydrodynamic fields in more detail in Section
\ref{sec:fusrl}.
\subsection{Summary of the ``Physical" Approach}
One should to realize that the consideration of anomalous scaling in this
Section is just an attempt to explain in terms of cascades, eddies and
their interaction the complicated multi-step processes in a system with
strong interactions without a small parameter.  It has not been a simple
task because our ``physical" intuition is usually restricted to the first
or second order of the perturbation approach while the origin of strong
renormalization of scaling behavior is related to taking into account an
infinite series of terms in perturbation theory.  In result our
``physical" approach of this Section cannot be considered as consistent.
For example in the beginning of our ``explanation" we discussed some group
of eddies with well separated scales $R\gg x \gg y \gg \eta$. Then it was
allowed to average our fields in a few consequent steps, over
$\eta$-ensemble, then over $y$-, $x$-, and finally over $R$-ensemble and
to make use of expansions of the fields in powers of the gradients of the
velocity fields, taking into account just the first non-trivial term.
Then the main contribution to the renormalization of the scaling came from
the multi-step processes in which our "group of eddies" are packaged
densely in the inertial range of scales. Therefore the applicability
parameter of this ``physical way of thinking" which should to be small is
really of the order of unity. This does not change qualitatively the
 physical picture of the anomalous behavior but leads to some additional
 contribution to the correlation functions which may be important.  One
direct way to find all of these contributions and to develop consistent
description of anomalous scaling in hydrodynamic turbulence is to make use
of the diagrammatic perturbation approach. That is the topic of the next
Section.

\section{Diagrammatic analysis}
\label{sec:diagr}
The theory of turbulence is a theory of strongly fluctuating hydrodynamic
motions. Systems with strong fluctuations are examined in quantum field
theory (where quantum fluctuations are relevant) and also in condensed
matter physics, e.g., in treating second order phase transitions (where
classical thermal fluctuations are relevant). Note that in both quantum
field theory and the theory of second order phase transitions a scaling
behavior of correlation functions (like in turbulence) is observed.  It is
known from these theories that adequate tools of theoretical investigation
of strong fluctuating systems are based upon functional integration (path
integration) methods, on different versions of the diagrammatic technique
and on related methods like renormalization-group analysis. Therefore a
consistent theory of turbulence should also be constructed in these terms.
In this section we demonstrate how the diagrammatic analysis leads to the
appearance of anomalous dimensions in the theory of turbulence.
\subsection{Scaling and Divergences}
The diagrammatic  technique for the problem  of turbulence was first
developed by Wyld \cite{W961}, who started from the Navier-Stokes equation
with a pumping force on the right-hand side. The Wyld technique enables
one to represent any correlation function characterizing the turbulent
flow as a series over the nonlinear interaction.  Unfortunately infrared
divergences appear in the technique.  Physically they are related to the
sweeping of small eddies by the velocity of the largest eddies. However
the sweeping does not change the energy of small eddies and consequently
does not influence the energy cascade. This means that the form of the
simultaneous correlation functions of the velocity is not sensitive to
sweeping.  In order to see this fact in each order of   diagrammatic
expansion we make use of the Belinicher-L'vov resummation\cite{BL87} which
corresponds to the strongly nonlinear change of variables: from the
Eulerian velocity to the quasi-Lagrangian ones (\ref{I1},\ref{I2}).  It
was demonstrated that after this resummation the infrared divergences in
the diagrammatic technique are absent and consequently the external scale
of the turbulence disappears from the diagrammatic equations. Therefore a
scale-invariant solution of these equations can exist.

An equation for the quasi-Lagrangian velocity {\bf u} which was introduced
by Eqs. (\ref{I1},\ref{I2}) can be derived from the Navier-Stokes
equation; it is
\begin{eqnarray}
\partial u_\alpha/\partial t &+&
\nabla_\beta\Bigl((u_\beta-u_{0\beta}) (u_\alpha-u_{0\alpha})\Bigr)+
\nabla_\alpha \tilde P
\nonumber \\
&=& \nu \nabla^2 u_\alpha +\tilde f_\alpha \,,
\quad \nabla_\alpha u_\alpha =0 \,.
\label{di01}
\end{eqnarray}
Here $\nu$ is the viscosity, $\tilde P$ and $\tilde{\bf f}$ are qL
variables related to the pressure $P$ and to the pumping force {\bf f} as
the qL velocity {\bf u} is related to {\bf v}, and the quantity
$u_{0\alpha}$ in (\ref{di01}) denotes $u_{\alpha}(t,{\bf r}_0)$.  The
equation (\ref{di01}) differs from the Navier-Stokes equation in the terms
${\bf u}_0$ subtracting the sweeping at the marked point ${\bf r}_0$. Note
that the presence of the marked point ${\bf r}_0$ in the theory means the
loss of homogeneity: this is the price for eliminating the infrared
divergencies from the diagrammatic expansion.  To find correlation
functions of the Eulerian velocity {\bf v} we should express these
correlation functions via correlation functions of the qL velocity {\bf
u}, which is not a trivial task.  Fortunately the simultaneous correlation
functions of the Eulerian velocity {\bf v} and of the qL velocity {\bf u}
coincide. Note that time-dependent correlation functions in these sets of
variables are different: moreover unlike the qL correlation functions the
time dependence of the correlation functions of the Eulerian velocity {\bf
v} is not of scaling type.

The Wyld diagrammatic expansion is formulated in terms of propagators and
vertices determined by the nonlinear term of the equation (\ref{di01}).
The propagators are the Green's function $G$ and the pair correlation
function $F$ of the velocities. We will treat these propagators for qL
variables. The $G$-function is the linear susceptibility determining the
average $\langle u_\alpha\rangle$ which arises as a response to the
nonzero average $\langle\tilde f_\alpha\rangle$:
\begin{equation}
G_{\alpha\beta}(t,{\bf r}_1,{\bf r}_2)=
-i\delta \langle u_\alpha(t,{\bf r}_1)\rangle/
\delta \langle\tilde f_\beta(0,{\bf r}_2)\rangle
\label{di02} \,,
\end{equation}
where $\tilde{f}_\alpha$ is the pumping force in the right-hand side of
(\ref{di01}). The $F$-function is the pair correlation function of
qL velocities:
\begin{equation}
F_{\alpha\beta}(t,{\bf r}_1,{\bf r}_2)=
\langle u_\alpha(t,{\bf r}_1)
u_\beta(0,{\bf r}_2) \rangle
\label{di03} \,.
\end{equation}
Note that the propagators $G$ and $F$ in qL variables depend separately on
coordinates of the points ${\bf r}_1$, ${\bf r}_2$ which is the
consequence of the loss of homogeneity. However the simultaneous
correlation function $F(t=0,{\bf r}_1,{\bf r}_2)$ depends only on the
difference ${\bf r}_1-{\bf r}_2$ since it should coincide with the
simultaneous correlation function of Eulerian velocities.

We expect that the propagators (\ref{di02},\ref{di03}) possess a scaling
behavior. To establish the character of this behavior we can utilize the
KO-41 dimensional estimations. For the pair correlation function
(\ref{di03}) we obtain from the estimate (\ref{veloc}) for the
characteristic velocity
\begin{equation}
F(t,{\bf r}_1,{\bf r}_2)\sim(\bar\varepsilon R)^{2/3}
\label{di04} \,,
\end{equation}
where $R$ is the characteristic scale. For the simultaneous correlation
function $R$ coincides with the separation between the points ${\bf r}_1$
and ${\bf r}_2$, and for nonzero $t$ it is determined also by the
differences ${\bf r}_1-{\bf r}_0$ and ${\bf r}_2-{\bf r}_0$, where ${\bf
r}_0$ is the marked point. The characteristic time of $F$ can also be
estimated. It coincides with the turnover time $\tau_{_R}\sim (R^2/ \bar
\varepsilon)^{1/3}$ (it is so only in qL variables).  The scaling behavior
of the Green's function can be established if one supposes that
fluctuations of {\bf u} does not drastically change the character of the
response to the external force. Then
\begin{equation}
G(t,{\bf r}_1,{\bf r}_2)\sim R^{-3}
\label{di05} \,,
\end{equation}
where $R$ is as before the characteristic scale. The question arises:  can
such scaling behavior be obtained as a solution of diagrammatic equations?

To answer this question we should first reformulate the diagram technique
in terms of the bare vertices but with the dressed propagators $F$ and
$G$.  Then one can easily check that the scaling behavior of $F$ and $G$
determined by the estimates (\ref{di04},\ref{di05}) is reproduced in any
order of the perturbation theory. But this is not sufficient to justify
the assertion that $F$ and $G$ actually possess such scaling behavior. The
reason for this was long ago recognized in the theory of second order
phase transitions. Reformulating the diagrammatic series for the
correlation functions of the order parameter in terms of the bare
interaction vertex but with the dressed pair correlation function with its
suitable scaling exponent, one can check that this exponent is reproduced
in each order of the perturbation theory. Besides one immediately
encounters logarithmic ultraviolet divergences which arise in each order
of the perturbation expansion. After summing, the logarithmic corrections
generate power factors which strongly renormalize the na\"{\i}ve
exponents.

Fortunately this phenomenon does not occur in the theory of turbulence.
As was demonstrated by Belinicher and L'vov \cite{BL87} in qL variables
there are neither  infrared nor ultraviolet divergences in the
diagrammatic expansion for $G$ and $F$, if (\ref{di04},\ref{di05}) are
assumed. The analogous theorem can be proved for high order correlation
functions of {\bf u} and nonlinear susceptibilities: the KO-41 exponents
are reproduced in the diagrammatic series and both ultraviolet and
infrared divergences are absent. This property is the ground for the
assertion that in the consistent theory the simultaneous correlation
functions actually have na\"{\i}ve KO-41 exponents.  To be more precise we
should note that this property is true for the correlation functions of
{\bf u} in {\bf k}-representation. In passing to {\bf r}-representation we
encounter infrared divergences related to the fact that the average value
$\langle{\bf u}^2\rangle$ (coinciding with $\langle{\bf v}^2\rangle$) is
determined by the largest scale (the scale of pumping) and is consequently
given by a formally diverging integral. To avoid this difficulty we should
consider the correlation functions of such objects as velocity
differences, in expressions for which the infrared divergences cancel.

The next question is: does the absence of ultraviolet divergences in
diagrams for correlation functions of {\bf u} mean that all objects in the
theory of turbulence can be characterized by KO-41 exponents?  Our answer
is ``no", since ultraviolet divergences can be observed for more
complicated objects. Namely, we will demonstrate that the ultraviolet
logarithms immediately arise in the diagrams for correlation functions of
powers of the velocity gradient $\nabla_\alpha v_\beta$. The simplest
example of such a correlation function is $K_{\varepsilon\varepsilon}({\bf
r}_1,{\bf r}_2)= \langle\langle\varepsilon({\bf r}_1)\varepsilon({\bf
r}_2)\rangle\rangle$ since the energy dissipation rate $\varepsilon$ is
proportional to the second power of $\nabla_\alpha v_\beta$.  Those
logarithms as in the theory of second order phase transitions lead to the
renormalization of the scaling behavior with respect to the normal KO-41
one that is anomalous scaling.
\subsection{Ultraviolet logarithms}
Let us analyze the diagrammatic expansion for the correlation function
$K_{\varepsilon\varepsilon}$ in qL variables in terms of the dressed
propagators $F$ and $G$. The first diagram for $K_{\varepsilon
\varepsilon}({\bf r}_1,{\bf r}_2)$ is depicted in Fig.~\ref{fig:fig1},
where a wavy line corresponds to the pair correlation function
(\ref{di03}) and an empty circle marks the point ${\bf r}_1$ or ${\bf
r}_2$. These points are designated by $1$ and $2$ in Fig.~\ref{fig:fig1}.
Since $\varepsilon$ is proportional to the second power of $\nabla_\alpha
v_\beta$ the gradients should be taken of both propagators attached to an
empty circle. Using the estimate (\ref{di04}) for the function $F$ we
immediately obtain that the diagram depicted in Fig.~\ref{fig:fig1} gives
the expression possessing the normal KO-41 behavior $\propto R^{-8/3}$.
This diagram reproduces Golitsin's result \cite{G962}.  No any divergence
is produced by this diagram.  To observe divergences we need to examine
higher order diagrams.

\begin{figure}

\setlength{\unitlength}{.2cm}
\begin{picture}(17,16)(-12,0)
  \put(1,7.5){$1$}
  \put(2.5,8){\circle{1}}
\multiput(4,8)(2,2){3}{\oval(2,2)[t,l]}
\multiput(4,10)(2,2){2}{\oval(2,2)[b,r]}
\multiput(4,8)(2,-2){3}{\oval(2,2)[b,l]}
\multiput(4,6)(2,-2){2}{\oval(2,2)[t,r]}
  \put(13.5,8){\circle{1}}
  \put(14.5,7.5){$2$}
\multiput(12,8)(-2,2){3}{\oval(2,2)[t,r]}
\multiput(12,10)(-2,2){2}{\oval(2,2)[b,l]}
\multiput(12,8)(-2,-2){3}{\oval(2,2)[b,r]}
\multiput(12,6)(-2,-2){2}{\oval(2,2)[t,l]}
  \put(7.5,7.5){$R$}
\end{picture}
\caption{The first diagram for $K_{\varepsilon\varepsilon}$
producing the normal KO-41 scaling.}
\label{fig:fig1}
\end{figure}
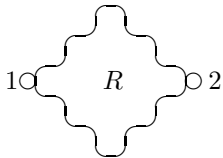

\begin{figure}
\setlength{\unitlength}{.15cm}
\begin{picture}(60,16)(3,0)
  \put(2.5,2){\circle{1}}
\multiput(3.5,2)(2,0){6}{\oval(1,.8)[t]}
\multiput(4.5,2)(2,0){6}{\oval(1,.8)[b]}
  \put(15,2){\circle*{.5}}
\put(15,2){\line(1,0){6}}
\multiput(21.5,2)(2,0){3}{\oval(1,.8)[t]}
\multiput(22.5,2)(2,0){3}{\oval(1,.8)[b]}
  \put(27.5,2){\circle{1}}
\multiput(15,2.5)(0,2){6}{\oval(.8,1)[l]}
\multiput(15,3.5)(0,2){6}{\oval(.8,1)[r]}
  \put(15,14){\circle*{.5}}
\multiput(4,2)(2,2){6}{\oval(2,2)[t,l]}
\multiput(4,4)(2,2){6}{\oval(2,2)[b,r]}
  \put(15,14){\line(1,-1){6}}
\multiput(21,7)(2,-2){3}{\oval(2,2)[t,r]}
\multiput(23,7)(2,-2){3}{\oval(2,2)[b,l]}
\put(18,5){$r$}
  \put(32.5,2){\circle{1}}
\multiput(33.5,2)(2,0){6}{\oval(1,.8)[t]}
\multiput(34.5,2)(2,0){6}{\oval(1,.8)[b]}
  \put(45,2){\circle*{.5}}
\put(45,2){\line(1,0){6}}
\multiput(51.5,2)(2,0){3}{\oval(1,.8)[t]}
\multiput(52.5,2)(2,0){3}{\oval(1,.8)[b]}
  \put(57.5,2){\circle{1}}
\multiput(45,2.5)(0,2){3}{\oval(.8,1)[l]}
\multiput(45,3.5)(0,2){3}{\oval(.8,1)[r]}
\put(45,8){\line(0,1){6}}
  \put(45,14){\circle*{.5}}
\multiput(34,2)(2,2){6}{\oval(2,2)[t,l]}
\multiput(34,4)(2,2){6}{\oval(2,2)[b,r]}
\multiput(45,13)(2,-2){6}{\oval(2,2)[t,r]}
\multiput(47,13)(2,-2){6}{\oval(2,2)[b,l]}
\put(48,5){$r$}

\end{picture}

\caption{The first diagrams for $K_{\varepsilon\varepsilon}$ producing
ultraviolet logarithms.}
\label{fig:fig2}
\end{figure}
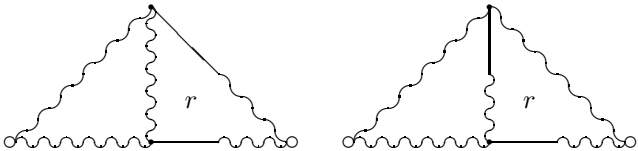

The diagrams of the next order are depicted in Fig.~\ref{fig:fig2}, where
the new combined wavy-straight lines designate the Green's function
(\ref{di02}): The wavy part corresponds to the variation of the velocity
{\bf u} and the straight part corresponds to the variation of the force
{\bf f}. The vertices on the diagrams are determined by the nonlinear term
in the equation (\ref{di01}): there are two wavy and one straight line
attached to each vertex. Two extra two-loop diagrams should also be taken
into account which can be obtained from the diagrams reproduced in
Fig.~\ref{fig:fig2} by converting left and right sides.  Using the
estimates (\ref{di04},\ref{di05}) we find that all of these diagrams give
us expressions for $K_{\varepsilon\varepsilon}$ which behave as
$R^{-8/3}$. However there are also logarithmic divergences in these
diagrams related to the loops marked by the letter ``$r$''. The lines of
the loop of the first diagram correspond to the product $\nabla G \nabla G
F$. In principle there are also $\nabla$'s in the vertices but the
structure of the qL vertex determined by (\ref{di01}) is such that it
produces the gradient of the attached propagator with the smallest wave
vector \cite{BL87,L991}, that is with the largest characteristic scale.
This means that for loops with separations $r$ (see Fig.~\ref{fig:fig2})
which are smaller than the separation $R$ between the points ${\bf r}_1$
and ${\bf r}_2$, the gradients producing by the vertices are external to
the loops. Thus the expression corresponding to the loop can be obtained
by integration of $\nabla G \nabla G F$ over the two times and coordinates
of two points corresponding to the vertices of the diagram. Using the
estimates (\ref{di04},\ref{di05}) we conclude that this integration is
dimensionless and produces consequently a logarithm. The upper limit in
this logarithmic integration is $R$ and the lower limit is the viscous
length $\eta$ at which the estimates (\ref{di04},\ref{di05}) fail,
therefore the logarithm produced by the right loop is $\ln(R/\eta)$. Then
the left loop gives $R^{-8/3}$ and the final contribution of the diagram
is $\propto R^{-8/3}\ln(R/\eta)$.  The same is correct for other diagrams
of the same order.

\begin{figure}

\setlength{\unitlength}{.2cm}
\begin{picture}(60,16)(3,0)

  \put(2.5,2){\circle{1}}
\multiput(3.5,2)(2,0){6}{\oval(1,.8)[t]}
\multiput(4.5,2)(2,0){6}{\oval(1,.8)[b]}
  \put(15,2){\circle*{.5}}
\put(15,2){\line(1,0){6}}
\multiput(21.5,2)(2,0){3}{\oval(1,.8)[t]}
\multiput(22.5,2)(2,0){3}{\oval(1,.8)[b]}
\put(15,14){\line(1,0){6}}
\multiput(21.5,14)(2,0){3}{\oval(1,.8)[t]}
\multiput(22.5,14)(2,0){3}{\oval(1,.8)[b]}
  \put(39.5,2){\circle{1}}
\multiput(15,2.5)(0,2){6}{\oval(.8,1)[l]}
\multiput(15,3.5)(0,2){6}{\oval(.8,1)[r]}
\multiput(27,2.5)(0,2){6}{\oval(.8,1)[l]}
\multiput(27,3.5)(0,2){6}{\oval(.8,1)[r]}
  \put(15,14){\circle*{.5}}
\multiput(4,2)(2,2){6}{\oval(2,2)[t,l]}
\multiput(4,4)(2,2){6}{\oval(2,2)[b,r]}
\put(27,2){\line(1,0){6}}
\multiput(33.5,2)(2,0){3}{\oval(1,.8)[t]}
\multiput(34.5,2)(2,0){3}{\oval(1,.8)[b]}
  \put(27,14){\circle*{.5}}
  \put(27,2){\circle*{.5}}
  \put(27,14){\line(1,-1){6}}
\multiput(33,7)(2,-2){3}{\oval(2,2)[t,r]}
\multiput(35,7)(2,-2){3}{\oval(2,2)[b,l]}
  \put(20,7){$R_2$}
  \put(30,5){$R_1$}

\end{picture}

\caption{The two-loop diagram for $K_{\varepsilon\varepsilon}$
producing the second power of the ultraviolet logarithm.}
\label{fig:fig3}
\end{figure}
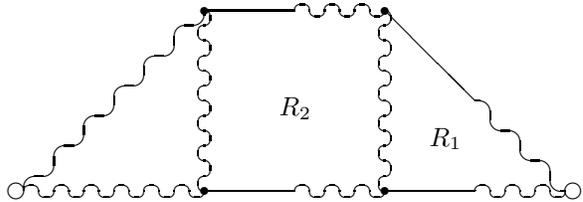

\begin{figure}

\setlength{\unitlength}{.2cm}
\begin{picture}(60,16)(3,0)

  \put(2.5,2){\circle{1}}
\multiput(3.5,2)(2,0){6}{\oval(1,.8)[t]}
\multiput(4.5,2)(2,0){6}{\oval(1,.8)[b]}
  \put(15,2){\circle*{.5}}
\put(15,2){\line(1,0){6}}
\multiput(21.5,2)(2,0){3}{\oval(1,.8)[t]}
\multiput(22.5,2)(2,0){3}{\oval(1,.8)[b]}
\put(15,14){\line(1,0){6}}
\multiput(21.5,14)(2,0){3}{\oval(1,.8)[t]}
\multiput(22.5,14)(2,0){3}{\oval(1,.8)[b]}
  \put(39.5,2){\circle{1}}
\multiput(16,2)(2,2){6}{\oval(2,2)[t,l]}
\multiput(16,4)(2,2){6}{\oval(2,2)[b,r]}
\multiput(15,13)(2,-2){6}{\oval(2,2)[t,r]}
\multiput(17,13)(2,-2){6}{\oval(2,2)[b,l]}
  \put(15,14){\circle*{.5}}
\multiput(4,2)(2,2){6}{\oval(2,2)[t,l]}
\multiput(4,4)(2,2){6}{\oval(2,2)[b,r]}
\put(27,2){\line(1,0){6}}
\multiput(33.5,2)(2,0){3}{\oval(1,.8)[t]}
\multiput(34.5,2)(2,0){3}{\oval(1,.8)[b]}
  \put(27,14){\circle*{.5}}
  \put(27,2){\circle*{.5}}
  \put(27,14){\line(1,-1){6}}
\multiput(33,7)(2,-2){3}{\oval(2,2)[t,r]}
\multiput(35,7)(2,-2){3}{\oval(2,2)[b,l]}

\end{picture}

\caption{The two-loop diagram for $K_{\varepsilon\varepsilon}$
producing only the first power of the ultraviolet logarithm.}
\label{fig:fig4}
\end{figure}
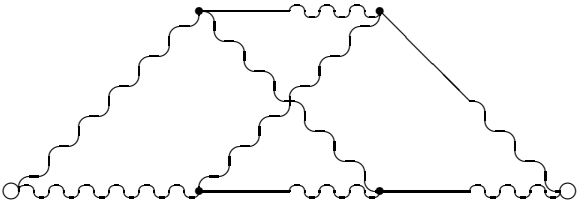

Let us now consider higher order diagrams for
$K_{\varepsilon\varepsilon}$.  An example is presented in
Fig.~\ref{fig:fig3} where a three-loop diagram is depicted. Based on the
above analysis one can readily recognize that this diagram gives the
second power of the ultraviolet logarithm.  The first logarithm is
produced by the right loop (marked by ``$R_1$''). It originates in the
integration over separations $R_1$ between the viscous scale $\eta$ and
the characteristic separation $R_2$ of the middle loop, this logarithm
being $\ln(R_2/\eta)$. The middle loop produces an extra logarithmic
integration over $d\ln(R_2/\eta)$. The result of this integration is $\int
d\ln(R_2/ \eta) \ln(R_2/\eta)$$ =(1/2)\ln^2(R/\eta)$ where $R$ is the
separation between the points ${\bf r}_1$ and ${\bf r}_2$.  Thus we
actually find the second power of the logarithm, arising as a prefactor of
$R^{-8/3}$ produced by the left loop.  Note that some three-loop diagrams
will not produce the second power of the logarithm.  An example of such a
diagram is given in Fig.~\ref{fig:fig4}. This diagram contains crossed
lines. It prohibits the existence of the region of integration $R\gg
R_2\gg R_1$ producing the second power of the logarithm.

Generalizing the above analysis we can conclude that diagrams of the
$n$-th order will produce the normal KO-41 factor $R^{-8/3}$ with
prefactors which are different powers of the logarithm $\ln(R/\eta)$ up to
the $n$-th power. Thus we encounter a series over $\ln(R/\eta)$, which
could be an arbitrary function of $\ln(R/\eta)$. In the next subsection we
will argue that this function is an exponential one, which is a power of
$R/\eta$. Such a function in the prefactor leads to the substitution of
the normal KO-41 dimensionality by other ones: this is the mechanism
producing an anomalous dimension.

Let us recognize that the origin of the ultraviolet logarithms (leading to
anomalous dimensions) in the diagrammatic series is related to the fact
that instead of the correlation functions of the velocities we have taken
the correlation function of the powers of the velocity gradients. Indeed,
we have seen that the logarithms are produced by loops with additional
differentiation of propagators. In our example these additional
differentiation were related to the structure of $\varepsilon$ which is
proportional to the second power of the velocity gradients. The loops
without additional differentiations do not produce ultraviolet logarithms
due to the main property of the qL vertex: the differentiation at the
vertex is taken of the propagator with the largest characteristic scale.
Therefore we should differentiate the propagators which are always outside
a ``ultraviolet loop'' (that is a loop with a small separation). The same
property of the qL vertex supplies the ``reproduction'' of extra
differentiations: we have seen that the differentiations external to the
first ``ultraviolet loop'' were internal for the next ``ultraviolet
loop'',  and so on, and so forth. A repetition of such loops results in
the accumulation of powers of the logarithms.  It is clear that the same
mechanism will work for any correlation function of objects containing
velocity gradients (but at least two gradients).  Therefore we expect that
all such objects should possess anomalous dimensions.

\subsection{Anomalous Dimensions}
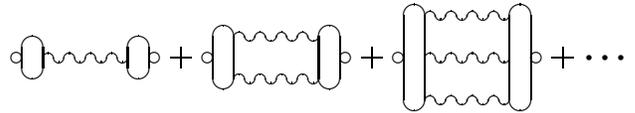
\begin{figure}

\setlength{\unitlength}{.141cm}
\begin{picture}(62,14)(3,0)
  \put(2.5,7){\circle{1}}
  \put(4,7){\oval(2,4)}
\multiput(5.5,7)(2,0){4}{\oval(1,.8)[t]}
\multiput(6.5,7)(2,0){4}{\oval(1,.8)[b]}
  \put(14,7){\oval(2,4)}
  \put(15.5,7){\circle{1}}
\put(17,7){\line(1,0){2}}
\put(18,6){\line(0,1){2}}
  \put(20.5,7){\circle{1}}
  \put(22,7){\oval(2,6)}
\multiput(23.5,5)(2,0){4}{\oval(1,.8)[t]}
\multiput(24.5,5)(2,0){4}{\oval(1,.8)[b]}
\multiput(23.5,9)(2,0){4}{\oval(1,.8)[t]}
\multiput(24.5,9)(2,0){4}{\oval(1,.8)[b]}
  \put(32,7){\oval(2,6)}
  \put(33.5,7){\circle{1}}
\put(35,7){\line(1,0){2}}
\put(36,6){\line(0,1){2}}
  \put(38.5,7){\circle{1}}
  \put(40,7){\oval(2,10)}
\multiput(41.5,7)(2,0){4}{\oval(1,.8)[t]}
\multiput(42.5,7)(2,0){4}{\oval(1,.8)[b]}
\multiput(41.5,11)(2,0){4}{\oval(1,.8)[t]}
\multiput(42.5,11)(2,0){4}{\oval(1,.8)[b]}
\multiput(41.5,3)(2,0){4}{\oval(1,.8)[t]}
\multiput(42.5,3)(2,0){4}{\oval(1,.8)[b]}
  \put(50,7){\oval(2,10)}
  \put(51.5,7){\circle{1}}
\put(53,7){\line(1,0){2}}
\put(54,6){\line(0,1){2}}
  \put(56.5,7){\circle*{.5}}
  \put(58,7){\circle*{.5}}
  \put(59.5,7){\circle*{.5}}

\end{picture}

\caption{The formally exact diagrammatic representation for
$K_{\varepsilon\varepsilon}$, the first terms of an infinite series.}
\label{fig:fig5}
\end{figure}

Let us return to the analysis of the correlation function $K_{\varepsilon
\varepsilon}$. In the framework of the Wyld technique a formally exact
diagram representation for the correlation function can be formulated
originating from the fact that in each diagram for $K_{\varepsilon
\varepsilon}$ there exists only one cut going along all $F$-functions.
This enables us to formulate the representation depicted in
Fig.~\ref{fig:fig5}, which is analogous to the representation of the
imaginary parts of propagators in quantum field theory.  Here we have
classified diagrams for $K_{\varepsilon\varepsilon}$ in accordance with
the number of $F$-functions in our marked cut (since this number runs from
$1$ to $\infty$ we obtain an infinite number of terms); the ovals
designate objects which are sums of the diagrams at the left and at the
right sides of the marked cut. As before empty circles mark the points
${\bf r}_1$ and ${\bf r}_2$ where $\varepsilon$ taken.

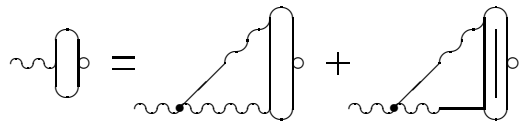
\begin{figure}

\setlength{\unitlength}{.15cm}
\begin{picture}(45,13)
\multiput(.5,6)(2,0){2}{\oval(1,.8)[t]}
\multiput(1.5,6)(2,0){2}{\oval(1,.8)[b]}
 \put(5,6){\oval(2,6)}
 \put(6.5,6){\circle{1}}
   \put(9,6.5){\line(1,0){2}}
   \put(9,5.5){\line(1,0){2}}
\multiput(11.5,2)(2,0){2}{\oval(1,.8)[t]}
\multiput(12.5,2)(2,0){2}{\oval(1,.8)[b]}
   \put(15,2){\circle*{.8}}
  \put(15,2){\line(1,1){4}}
  \multiput(20,6)(2,2){2}{\oval(2,2)[t,l]}
  \multiput(20,8)(2,2){2}{\oval(2,2)[b,r]}
 \multiput(15.5,2)(2,0){4}{\oval(1,.8)[t]}
 \multiput(16.5,2)(2,0){4}{\oval(1,.8)[b]}
\put(24,6){\oval(2,10)}
 \put(25.5,6){\circle{1}}
  \put(28,6){\line(1,0){2}}
  \put(29,5){\line(0,1){2}}
\multiput(30.5,2)(2,0){2}{\oval(1,.8)[t]}
\multiput(31.5,2)(2,0){2}{\oval(1,.8)[b]}
   \put(34,2){\circle*{.8}}
  \put(34,2){\line(1,1){4}}
  \multiput(39,6)(2,2){2}{\oval(2,2)[t,l]}
  \multiput(39,8)(2,2){2}{\oval(2,2)[b,r]}
 \multiput(34.5,2)(2,0){2}{\oval(1,.8)[t]}
 \multiput(35.5,2)(2,0){2}{\oval(1,.8)[b]}
 \put(38,2){\line(1,0){4}}
\put(43,6){\oval(2,10)}
\put(43,3){\line(0,1){6}}
 \put(44.5,6){\circle{1}}
\end{picture}

\caption{The diagrammatic representation for the block entering the
``one-bridge'' contribution to $K_{\varepsilon\varepsilon}$.}
\label{fig:fig6}

\end{figure}

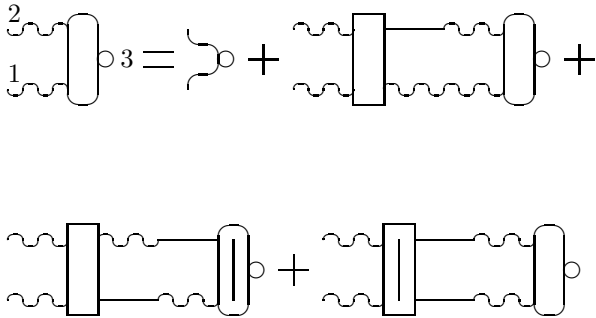
\begin{figure}

\setlength{\unitlength}{.2cm}
\begin{picture}(41,28)(1,-14)
\multiput(2.5,5)(2,0){2}{\oval(1,.8)[t]}
\multiput(1.5,5)(2,0){2}{\oval(1,.8)[b]}
  \put(1,5.5){$1$}
\multiput(2.5,9)(2,0){2}{\oval(1,.8)[t]}
\multiput(1.5,9)(2,0){2}{\oval(1,.8)[b]}
  \put(1,9.5){$2$}
\put(6,7){\oval(2,6)}
  \put(7.5,7){\circle{1}}
  \put(8.5,6.5){$3$}
  \put(10,7.5){\line(1,0){2}}
  \put(10,6.5){\line(1,0){2}}
\put(14,9){\oval(2,2)[l,b]}
\put(14,5){\oval(2,2)[l,t]}
\put(14,7){\oval(2,2)[r]}
\put(15.5,7){\circle{1}}
  \put(17,7){\line(1,0){2}}
  \put(18,6){\line(0,1){2}}
\multiput(20.5,5)(2,0){2}{\oval(1,.8)[t]}
\multiput(21.5,5)(2,0){2}{\oval(1,.8)[b]}
\multiput(20.5,9)(2,0){2}{\oval(1,.8)[t]}
\multiput(21.5,9)(2,0){2}{\oval(1,.8)[b]}
  \put(24,4){\framebox(2,6)}
\multiput(26.5,5)(2,0){4}{\oval(1,.8)[t]}
\multiput(27.5,5)(2,0){4}{\oval(1,.8)[b]}
\put(26,9){\line(1,0){4}}
\multiput(30.5,9)(2,0){2}{\oval(1,.8)[t]}
\multiput(31.5,9)(2,0){2}{\oval(1,.8)[b]}
  \put(35,7){\oval(2,6)}
  \put(36.5,7){\circle{1}}
  \put(38,7){\line(1,0){2}}
  \put(39,6){\line(0,1){2}}
\multiput(1.5,-5)(2,0){2}{\oval(1,.8)[t]}
\multiput(2.5,-5)(2,0){2}{\oval(1,.8)[b]}
\multiput(1.5,-9)(2,0){2}{\oval(1,.8)[t]}
\multiput(2.5,-9)(2,0){2}{\oval(1,.8)[b]}
  \put(5,-10){\framebox(2,6)}
\put(7,-9){\line(1,0){4}}
\multiput(11.5,-9)(2,0){2}{\oval(1,.8)[t]}
\multiput(12.5,-9)(2,0){2}{\oval(1,.8)[b]}
\put(11,-5){\line(1,0){4}}
\multiput(7.5,-5)(2,0){2}{\oval(1,.8)[t]}
\multiput(8.5,-5)(2,0){2}{\oval(1,.8)[b]}
  \put(16,-7){\oval(2,6)}
  \put(16,-5){\line(0,-1){4}}
  \put(17.5,-7){\circle{1}}
  \put(19,-7){\line(1,0){2}}
  \put(20,-8){\line(0,1){2}}
\multiput(22.5,-5)(2,0){2}{\oval(1,.8)[t]}
\multiput(23.5,-5)(2,0){2}{\oval(1,.8)[b]}
\multiput(22.5,-9)(2,0){2}{\oval(1,.8)[t]}
\multiput(23.5,-9)(2,0){2}{\oval(1,.8)[b]}
  \put(26,-10){\framebox(2,6)}
  \put(27,-9){\line(0,1){4}}
\put(28,-9){\line(1,0){4}}
\multiput(32.5,-9)(2,0){2}{\oval(1,.8)[t]}
\multiput(33.5,-9)(2,0){2}{\oval(1,.8)[b]}
\put(28,-5){\line(1,0){4}}
\multiput(32.5,-5)(2,0){2}{\oval(1,.8)[t]}
\multiput(33.5,-5)(2,0){2}{\oval(1,.8)[b]}
  \put(37,-7){\oval(2,6)}
  \put(38.5,-7){\circle{1}}
\end{picture}

\caption{The diagrammatic equation for the three-point object entering
the diagrammatic expression for $K_{\varepsilon\varepsilon}$.}
\label{fig:fig7}
\end{figure}

The first ``one-bridge'' term of the diagrammatic series depicted in
Fig.~\ref{fig:fig5} is determined by the block which can be represented as
the sum of two terms depicted in Fig.~\ref{fig:fig6}, where we have
introduced the new object designated by the oval with the inscribed line.
Unlike the empty oval the oval with the inscribed line has one straight
leg. Actually both ovals occurring in~Fig. \ref{fig:fig6} arise in the
diagrammatic equation for the oval entering the second ``two-bridge'' term
in Fig.~\ref{fig:fig5}.  Therefore we begin our analysis with the equation
for this oval. Diagrams for the oval can be classified according to the
possibility to cut the diagram along two lines. Note that those two lines
can correspond to two $G$-functions or to the functions $G$ and $F$ but
not to two $F$-functions since the cut with all $F$-functions is unique.
Designating by boxes the sums of the four-leg parts of the diagrams which
cannot be cut along two lines, we come after summation to the diagrammatic
equation presented in Fig.~\ref{fig:fig7}. The first term on the RHS here
designates the bare contribution determined by the double $\nabla$ in
$\varepsilon$. Other terms are related to fluctuations.

Similar diagrammatic relations can be established also for the three-point
object designated by the oval with the inscribed line and for the
analogous three-point object with two straight legs.  The relations can be
considered as a closed system of equations for these three ovals. Of
course the boxes entering these equations are not known explicitly.
Nevertheless it is possible to establish estimates for the boxes.
Following the analysis given in \cite{BL87,L991} one can demonstrate that
there are neither ultraviolet nor infrared divergences in the diagrams for
the boxes. It means that they can be estimated by the first contributions.
The first contribution to the empty box is determined by the $G$-function
with two attached vertices and the first contribution to the box with the
inscribed line is determined by the $F$-function with two attached
vertices.

The diagrammatic equation represented in Fig.~\ref{fig:fig7} can be
rewritten in the analytical form. For this let us introduce the
designation $\Upsilon$ for the oval on the RHS of the diagram. Thus is a
three point object and therefore $\Upsilon$ is a function $\Upsilon_
{\alpha\beta}(t_1,{\bf r}_1,t_2,{\bf r}_2,t_3,{\bf r}_3)$, where the
arrangement of the points $1,2,3$ is shown in Fig.~\ref{fig:fig7}, and
$\varepsilon$ is taken at the point $3$. Then the equation represented in
Fig.~\ref{fig:fig7} takes the form
\begin{eqnarray}
&& \Upsilon(t_1,{\bf r}_1,t_2,{\bf r}_2,t,{\bf r})=
\Upsilon_0(t_1,{\bf r}_1,t_2,{\bf r}_2,t,{\bf r})+
\nonumber \\
&& \int dt_3 d^3r_3 dt_4 d^3r_4
B(t_1,{\bf r}_1,t_2,{\bf r}_2,t_3,{\bf r}_3,t_4,{\bf r}_4)\times
\nonumber \\
&& \Upsilon(t_3,{\bf r}_3,t_4,
{\bf r}_4,t,{\bf r})+\dots
\label{di06} \ .
\end{eqnarray}
Here for brevity we have omitted subscripts and an additional term
determined by the oval with the inscribed line. The quantity $\Upsilon_0$
in (\ref{di06}) is the bare value of $\Upsilon$:
\begin{equation}
\Upsilon_{0}=\nu
\delta(t_1-t)\delta(t_2-t)\nabla\delta({\bf r}_1-{\bf r})
\nabla\delta({\bf r}_2-{\bf r})
\label{di07} \,.
\end{equation}
The kernel $B$ in (\ref{di06}) corresponds to the sum of boxes in
Fig.~\ref{fig:fig7} with attached lines. This kernel can be estimated as
$F\nabla G \nabla G$, where the double $\nabla$ originates from the
vertices produced by (\ref{di01}). Utilizing the estimates
(\ref{di04},\ref{di05}) and also $t\sim(R^2/\bar\varepsilon)^{1/3}$ we
conclude that the integration in (\ref{di06}) is dimensionless.  It is
possible to rescale two other ovals to achieve dimensionless integrations
in terms designated by dots in (\ref{di06}) and in two analogous equations
for other ovals. Thus we come to homogeneous equations.

Of course this homogeneity will be destroyed on scales of the order of
$\eta$. Therefore the structure of a solution will be as follows: on
scales larger than $\eta$ the function $\Upsilon$ is a sum of power-like
terms characterized by scaling exponents which could be excluded from
solving the equation (\ref{di06}) without $\Upsilon_0$ and with the
homogeneous kernel. The number of these terms is infinite.  This equation
is linear and consequently the coefficients of the powers cannot be
extracted from this equation. Actually the coefficients can be found from
the whole solution, with $\Upsilon_0$ taken into account and with account
of the solution on scales of the order of $\eta$.  Also the set of scaling
exponents does not depend on the details of the small-scale behavior of
the system since it can be extracted by solving the equation with the
kernel determined by the dynamics in the inertial interval.  Returning now
to $K_{\varepsilon\varepsilon}$ we conclude that the first two terms of
the expansion presented in Fig.~\ref{fig:fig5} produce a complicated
scaling behavior of this function: on scales larger than $\eta$
$K_{\varepsilon\varepsilon}$ is a sum of an infinite number of power-like
terms with the universal set of scaling exponents but with coefficients
sensitive to the dynamics on scales of the order of $\eta$.  All of these
terms possess anomalous dimensions related to the scaling behavior of
$\Upsilon$. The same assertion can also be proven for higher order terms
of this expansion, since for all ovals equations of the type of
(\ref{di06}) can be formulated leading to the same conclusions for the
scaling behavior.

Actually the situation is the same when we consider more complicated
correlation functions of fields constructed from velocity gradients since
the gradients will produce logarithms and consequently anomalous
dimensions. Any correlation function of such fields is as above a sum of
terms with different anomalous dimensions.  Therefore a question of great
interest is the calculation of the set of anomalous dimensions.
Unfortunately this is impossible to do analytically as we deal with a
theory where there is no small parameter. This means for example that we
cannot find an explicit expression for the kernel $B$ in (\ref{di06}) (it
is possible to do for the simple model considered in \cite{LPF4}).
Therefore the set of anomalous dimensions should be extracted from
experiment or numerics.  Nevertheless we can establish a number of
selection rules based on symmetry reasoning. These  rules allow us to
establish a set of relations between different correlation functions of
velocity gradients and to predict some intermediate asymptotics of the
correlation functions of the velocity differences. These rules will be
discussed in the next section.

\section{Fusion Rules and Operator Algebra in Turbulence}
\label{sec:fusrl}

Here we are going to formulate for hydrodynamic turbulence a set of {\it
fusion rules for fluctuating fields} as introduced by Polyakov
\cite{P969}, who investigated the scaling behavior of the correlation
functions of the order parameter near a second order phase transition
point. We think that Polyakov's idea (which is actually a kind of
multipole expansion) can be successfully applied to any system with strong
fluctuations where a scaling behavior is observed. Our treatment will be
particularly based on the results obtained in the preceding sections but
the fusion rules themselves can be constructed practically independently
starting from symmetry arguments. A convenient language for formulating
fusion rules is the so-called {\it operator algebra} introduced by
Kadanoff \cite{K969} and Wilson \cite{W969}. In this section we will
restrict ourselves to the analysis of simultaneous correlation functions
of the velocity gradient which coincide for both Eulerian and qL
velocities. These correlation functions, in contrast to different-time
correlation functions of qL velocity, possess space homogeneity which
essentially simplifies the analysis.

\subsection{Scaling Dimensions of Local Fields}

Here we will treat correlation functions of local fields $\varphi_j({\bf
r})$ constructed as different single-point products of the velocity
derivatives.  An example of a local field is the energy dissipation rate
$\varepsilon$ which is proportional to the second power of the velocity
gradient (\ref{dis}). We have seen in Section \ref{sec:diagr} that the
correlation function $K_{\varepsilon\varepsilon}=\langle\langle
\varepsilon\varepsilon\rangle\rangle$ contains an infinite number of terms
with different scaling behavior. The same is actually true for any
correlation function $\langle\langle \varphi_i({\bf
r})\varphi_j(0)\rangle\rangle$ of the local fields. To proceed in the
analysis of their scaling behavior it is worthwhile to extract a set of
local fields $A_n$ with  ``clean'' scaling behavior which are complicated
linear combinations of single-point products of the velocity derivatives.
The fields $\Psi_1$ and $\Psi_2$ introduced by (\ref{B8}) can be
considered as the first step in this direction.  Of course it is
impossible to construct explicitly the final expressions for $A_n$;
fortunately the expressions are not really needed for the analysis of the
scaling behavior of correlation functions in which we are interested.

Each local field $A_n$ by definition is characterized by its scaling
dimension $\Delta_n$ which means that the correlation function $\langle
A_n({\bf R}) A_n(0)\rangle \propto R^{-2\Delta_n}$. The question arises as
to the scaling behavior of the correlation function $\langle A_n({\bf R})
A_m(0)\rangle$. To establish this behavior one might recall, e.g., the
diagrammatic expansion for $K_{\varepsilon\varepsilon}$ presented in
Fig.~\ref{fig:fig5}.  The scaling behavior of the terms of this expansion
is characterized by the anomalous dimensions of the corresponding
diagrammatic objects designated by the ovals. The anomalous dimension of
each term is equal to the sum of the anomalous dimensions determined by
the pair of ovals. The same situation takes place for the correlation
function $\langle A_n({\bf R})A_m(0)\rangle$: in the diagrammatic language
it is the pair of diagrammatic ``fur-coats'' dressing the points {\bf R}
and $0$ and a number of ``bridges'' made from $F$-functions connecting
these ``fur-coats''. Each ``fur-coat'' has its own anomalous dimension and
therefore the anomalous dimension of $\langle A_n({\bf R})A_m(0)\rangle$
is equal to the sum of the anomalous dimensions of the ``fur-coats''
(since the $F$-function has no anomalous dimension). Thus we come to the
conclusion that the scaling index of the correlation function $\langle
A_n({\bf R})A_m(0)\rangle$ is equal to $\Delta_n+\Delta_m$. This means
that
\begin{equation}
\langle A_n({\bf R})A_m(0)\rangle
\propto R^{-\Delta_n-\Delta_m}
\label{co01} \,.
\end{equation}
This is the basic property of the fields $A_n$ which will be exploited
further.

Among the set $A_n$ of the local fields with definite scaling dimensions
one can extract a subset of the so-called primary fields $O_n$ which give
rise to all other fields $A_n$ by differentiation \cite{P974}. These
``field-descendants'' $A_n$ are usually referred as {\it secondary
fields}. The dimension $\Delta_n$ of any secondary field $A_n$ differs
from the dimension $\Delta_m$ of the corresponding primary field $O_m$ by
an integer number $l$:  $\Delta_n=\Delta_m+l$, the number $l$ being the
number of the differentiations needed to obtain $A_n$ from $O_m$. It is
obvious that the set $\Delta_m$ of the scaling dimensions of the primary
fields determine also the whole set $\Delta_n$.

Any local field $\varphi_j$ can be expanded in a  series over the fields
$A_n$ with some coefficients $\varphi_{j(n)}$:
\begin{equation}
\varphi_j({\bf r})=\sum_n\varphi_{j(n)} A_n({\bf r})
\label{co02} \,.
\end{equation}
This expansion enables one to reduce the correlation functions of the
field $\varphi_j$ to the correlation functions of the fields $A_n$.  It is
convenient to order the fields $A_n$ over the values of their scaling
dimensions:  $\Delta_1\le\Delta_2\le\Delta_3\dots$.  Unfortunately it is
impossible to find the values of $\Delta_n$ (to do this we should find
eigenfunctions of an equation of the  type of (\ref{di06}), the kernel of
which is not known explicitly) but it is possible to express the scaling
behavior of observable quantities in terms of $\Delta_n$. It is clear that
the principal scaling behavior of correlation functions of a local field
$\varphi_j$ is determined by the first nonzero term of its expansion into
a series over $A_n$. For example if the first terms of the expansions of
$\varphi_1$ and $\varphi_2$ are not equal to zero the scaling behavior of
the principal term in the correlation function $\langle \varphi_1({\bf
R})\varphi_2(0)\rangle$ is related to the correlation function $\langle
A_1 A_1 \rangle$ since in accordance with the definition all other
correlation functions $\langle A_m A_n \rangle$ decrease more rapidly than
$\langle A_1 A_1 \rangle$ as $R\to\infty$. Thus $\langle \varphi_1({\bf
R})\varphi_2(0)\rangle\propto R^{-2\Delta_1}$ as $R\to\infty$. Since the
dimensions of the secondary fields are larger then the dimensions of the
primary fields we first should be interested in the terms of the expansion
of $\varphi_j$ over $O_n$.

The introduction of the fields $A_n$ enables one not only to predict the
scaling behavior of the pair correlation functions but also to investigate
the asymptotic behavior of more complicated correlation functions.
Consider for instance a three-point correlation function of the fields
$\varphi_j$ which using (\ref{co02}) can be reduced to three-point
correlation functions of the fields $A_n$. Suppose now that among the
points there are two nearby points and the third point is separated from
the pair by a large distance $R$,  and we are interested in the behavior
of this correlation function as $R$ varies. It is obvious that in this
case the product of the fields $A_n({\bf r}_1)A_m({\bf r}_2)$ taken at the
nearby points behaves like a single-point object, which can be expanded
into a series over $A_n({\bf r})$, the point {\bf r} being located near
the points ${\bf r}_1$ and ${\bf r}_2$. Thus we come to the relations
\begin{equation}
A_{n}({\bf r}_1)A_{m}({\bf r}_2)=
\sum_l C_{mn,l}({\bf r}_1-{\bf r},{\bf r}_2-{\bf r})
A_{l}({\bf r})
\label{co05} \,,
\end{equation}
which are known as the operator algebra \cite{K969,W969}.  Actually it is
a kind of multipole expansion. Let us explain how the relation
(\ref{co05}) can be found explicitly. For this we should expand $A_n({\bf
r}_1)$ and $A_m({\bf r}_2)$ in a series over ${\bf r}_1-{\bf r}$ and ${\bf
r}_2-{\bf r}$. Then we arrive at products of the local fields taken at the
point {\bf r} which can be reexpanded into a series over $A_n({\bf r})$.

The relation (\ref{co05}) can be used in investigating any correlation
function of the fields $\varphi_j$ with two nearby points since it means
that the correlation functions of the products of the left-hand and of the
right-hand sides of this relation with any remote object coincide.  For
this investigation we should first expand the product $\varphi_j\varphi_i$
taken in nearby points in accordance with (\ref{co02}), then (\ref{co05})
enables one to present the result in the form of a series over $A_n$. Of
course this procedure is very complicated, but actually we can restrict
ourselves to the first terms of the expansion over $A_n$ which
considerably simplifies the procedure.

\subsection{Local Fields in Turbulence}

In this subsection we consider the local fields $\varphi_j$ built up from
the velocity derivatives $\nabla_\alpha v_\beta$ from the point of view of
symmetry. The symmetry imposes some restrictions on the mutual correlation
functions of various local fields.  For example the irreducible cross
correlation function of a scalar and of a pseudoscalar should be equal to
zero. The point is that the correlation function should change its sign
under inversion while it is impossible to construct such a pseudoscalar
correlation function from the separation vector {\bf R}. Let us stress
that different behavior of the fields $\varphi_j$ under time reversal
$t\to-t$ does not impose any restriction on the type of nonzero
correlation functions. This is because the time reversal symmetry in the
turbulent system is violated (due to the presence of a nonzero energy
flux).

Local fields with different transformation properties arise already in the
first order in velocity gradients, namely the stress tensor
$s_{\alpha\beta}=\case{1}{2}(\nabla_\alpha v_\beta+\nabla_\beta v_\alpha)$
and the vorticity ${\bbox\omega}={\bbox\nabla}\times{\bf v}$.  The number
of objects of the second order in $\nabla_\alpha v_\beta$ is larger:
there are two scalar fields $s^2$ and $\omega^2$, one pseudovector field
$s_{\alpha\beta} \omega_\beta$, three traceless symmetric tensor fields
$s_{\alpha \gamma} s_{\gamma\beta} -(1/3)s^2\delta_{\alpha\beta}$,
$\omega_\alpha \omega_\beta - (1/3)\omega^2\delta_{\alpha\beta}$ and
$\epsilon_{\alpha \beta\gamma}\omega_\beta s_{\gamma \delta}+
\epsilon_{\delta \beta\gamma} \omega_\beta s_{\gamma\alpha}$, one
irreducible pseudotensor of the third rank and one irreducible tensor of
the fourth rank.  It is obvious that the number of different fields which
can be built up from  $\nabla_\alpha v_\beta$ is infinite. A field
$\varphi_j$ with given transformation properties can be expanded into the
series (\ref{co02}) over the fields $A_n$ with the same transformation
properties:  a vector is expanded into a series over vectors ${\bf A}_n$,
an irreducible tensor of the second rank is expanded into a series over
tensors $A_{n,\alpha\beta}$ and so on, where the coefficients in these
expansions are scalars.

A special consideration is needed for the correlation functions of the
first powers of the strain tensor $s_{\alpha\beta}$ and the vorticity
${\bbox\omega}$. These correlation functions can be reduced to gradients
of the correlation function of the velocity. As we noted there are no
ultraviolet divergencies related to the velocity itself. This means that
both the strain tensor $s_{\alpha\beta}$ and the vorticity ${\bbox\omega}$
have the normal KO-41 scaling exponents $2/3$.  Additional restrictions
are related to the incompressibility condition $\nabla\cdot {\bf v}=0$.
For example, the cross correlation function of the velocity itself with
any scalar field $\varphi_j$ is equal to zero. To prove this, note that
the correlation function $\langle{\bf v}({\bf r}) \varphi_{j} (0) \rangle$
is a vector which can only  be directed along {\bf r}.  However we know
that the divergence of this vector should be equal to zero because of
incompressibility.  This is possible only if this vector is equal to zero
(the possibility $\langle{\bf v}({\bf r}) \varphi_{j}(0) \rangle \propto
{\bf r}/r^3$ is actually also excluded because of the singularity at ${\bf
r}=0$). This means that the correlation functions $\langle
s_{\alpha\beta}({\bf r})\varphi_{j} (0)\rangle$ and $\langle
{\bbox\omega}({\bf r})\varphi_{j}(0)\rangle$ are also zero since they can
be obtained from $\langle{\bf v}({\bf r})\varphi_{j}(0)\rangle$ by direct
differentiation.  Analogous reasons enable us to establish the form of the
correlation function $\langle\nabla_\gamma v_\alpha({\bf r}) A_{n, \beta}
(0) \rangle$ where $A_{n,\beta}$ is a vector field with the definite
scaling dimension $\Delta_{1,n}$. This correlation function should have
the scaling exponent $\Delta_{1,n}+2/3$ which gives two possible tensor
structures. Using now the property ${\bbox\nabla}\cdot{\bf v}=0$ we find:
\begin{eqnarray}
&& \langle \nabla_\gamma v_\alpha({\bf R})
A_{n,\beta}(0)\rangle\propto
\label{ll01} \\
&& \nabla_\gamma \left(
\left(\delta_{\alpha\beta}-\frac{1-3\Delta_{1,n}}{7-3\Delta_{1,n}}
\frac{R_\alpha R_\beta}{R^2}\right)
R^{1/3-\Delta_{1,n}}\right)
\nonumber \,.
\end{eqnarray}
We do not see any reason for the fields $s^2$ and $\omega^2$ to have
nonzero first coefficients in the expansion (\ref{co02}). It means, for
example, that the principal scaling behavior of all correlation functions
$\langle \langle s^2 s^2\rangle \rangle$, $\langle \langle \omega^2s^2
\rangle \rangle$ and $\langle \langle \omega^2 \omega^2 \rangle \rangle$
is the same: $\propto R^{-2\Delta_1}$ where $\Delta_1$ is the scaling
exponent of the field $O_1$ which is the scalar field with the smallest
exponent.  The same is true for scalar fields proportional to higher
powers of $\nabla_\alpha v_\beta$, say $(s^2)^n, n>1$, $(\omega^2)^n,
n>1$, $s^3=s_{\alpha \beta}s_{\beta\gamma}s_{\gamma \alpha}$,
$\omega_{\alpha} s_{\alpha \beta}\omega_{\beta}$, etc. Comparing the
principal scaling exponent $2\Delta_1$ with the exponent $8/3-2\Delta$
(where $\Delta$ is the anomalous dimension of $\varepsilon$ introduced in
Section \ref{sec:teles}) we conclude that $\Delta_1=4/3-\Delta$ or
$\Delta_1 =\mu/2$.  Besides the main terms $\propto O_1$ in the expansion
of a scalar field we can take into account the next term $\propto O_2$.
Such terms will produce corrections to the two-point correlation functions
of scalar fields which scale as $R^{-\Delta_1-\Delta_2}$.

Correlation functions of vector fields or of irreducible tensor fields
have scaling behavior different from that of scalar fields since they are
expanded into a series (\ref{co02}) over other fields $A_n$. For example
the principal term of $\langle \varphi_{i,\alpha\beta}({\bf R})
\varphi_{j,\alpha\beta}(0)\rangle$ behaves $\propto R^{-2\Delta_{2,1}}$,
where $\Delta_{2,1}$ is the scaling exponent of the 2-rank tensor field
$A_{1,\alpha\beta}$ with the smallest exponent.  Since the field
$s_{\alpha\beta}$ is among the terms of the expansion (\ref{co02}) a term
with a different field will be the leading one only if $\Delta_{2,1}<
2/3$.  In the opposite case the principal scaling behavior of the
correlation functions will be determined by the $s_{\alpha \beta}
$-proportional term which means $\Delta_{2,1}=2/3$. Thus we conclude that
$\Delta_{2, 1}\le 2/3$.  Similarly a correlation function of pseudovector
fields scales as $R^{-2\Delta_{1,1}^*}$ with the exponent
$\Delta_{1,1}^*\leq 2/3$ since the field ${\bbox\omega}$ is among the
fields $A_n$ in the expansion (\ref{co02}) for pseudovector fields. It is
possible to establish also a restriction for the correlation functions of
vector fields. The pair correlation functions of these fields scale as
$R^{-2\Delta_{1,1}}$  with the exponent $\Delta_{1,1}\leq 1+\Delta_1$
since $\nabla O_1$ is among the fields $A_n$ in the expansion (\ref{co02})
for vector fields.

Note that if the system possesses conformal symmetry then there exists a
set of strong selection rules for the coefficients in the right-hand side
of (\ref{co01}), established by Polyakov \cite{P974}.  Namely these
coefficients are non-zero for different values $\Delta_n$ and $\Delta_m$
only if these fields are secondary fields of the same primary field. This
is the consequence of the ``orthogonality rule'': the correlation
functions of different primary fields $O_n$ are equal to zero if the
system possesses conformal symmetry. This ``orthogonality rule'' could be
a basis for experimental checking: is fully developed hydrodynamic
turbulence conformal or not. This question arises in connection with the
recent work of Polyakov \cite{P993} who treated $2d$ turbulence in the
framework of the conformal approach (as is known \cite{BPZ4} for $2d$
systems conformal symmetry permits one to establish many properties of the
correlation functions, particularly possible sets of dimensions
$\Delta_n$).

One of the possible ways to check for conformal symmetry in turbulence is
to study experimentally the cross correlation function $\left\langle \left
\langle s^2({\bf R}) \varphi_{j\alpha\beta}(0) \right \rangle \right
\rangle$. Let us make the expansion (\ref{co02}) for both fields. In the
case of conformal symmetry the nonzero contribution to the correlation
function will produce only averages of the fields belonging to one
``family'' originating from some primary field. We expect that the main
contribution will be produced by an average of the form $\langle O_1
(3\nabla_\alpha \nabla_\beta O_1-\delta_{\alpha\beta}\nabla^2 O_1)\rangle$
which gives
\begin{equation}
\left\langle \left\langle
s^2({\bf R}) \varphi_{j\alpha\beta}(0)
\right\rangle\right\rangle
\propto R^{-(2+2\Delta_{2,1})}
\label{bb12} \,.
\end{equation}
If conformal symmetry does not hold the exponent in (\ref{bb12}) will not
coincide with $2+2\Delta_{2,1}$.

\subsection{Fusion Rules for Velocity Differences}

Here we will examine  the asymptotic behavior of correlation functions of
the velocity differences. Consider the case when there are two sets of
nearby points separated by a large length $R$ and examine the behavior of
different correlation functions at varying $R$. We will demonstrate that
in such a procedure the anomalous dimensions should be revealed. Of course
our treatment is restricted to scales belonging to the inertial subrange.

In the preceding subsection we have considered local fields $\varphi_j$
built up from the velocity derivatives, since correlation functions of
these fields have no infrared contributions. Actually this property
enabled us to develop the expansion (\ref{co02}) into a series over the
fields $A_j$ with definite scaling dimensions.  Here we wish to note that
the behavior of the correlation functions of the velocity differences can
be extracted from one of the correlation functions of the velocity
derivatives using the simple relation
\begin{equation}
v_\alpha({\bf r}_1)-v_\alpha({\bf r}_2)=
\int\limits_2^1 dr_\beta \nabla_\beta v_\alpha
\label{co03} \,,
\end{equation}
where the integral is taken along any curve connecting the point $1$
to the point $2$.

The operator algebra (\ref{co05}) together with (\ref{co03}) enables us to
reduce any product of velocity differences taken at nearby points to a
single-point object. Consider as an example the second power of the
velocity difference $({\bf v}_1-{\bf v}_2)^2$. Using
(\ref{co05},\ref{co03}) we can ``fuse'' this object into a single point
{\bf r} at which  the expansion over all  fields is generated.  It is
natural to expect that the principal term in this expansion is determined
by $O_1$:
\begin{equation}
({\bf v}_1-{\bf v}_2)^2 \to
f^{(1)}(r_{12})O_1(({\bf r}_1+{\bf r}_2)/2)\,,
\label{co33} \,,
\end{equation}
where {\bf r} is chosen to be equal to $({\bf r}_1+{\bf r}_2)/2$
and $r_{12}=|{\bf r}_1-{\bf r}_2|$. It means, for example, that
\begin{equation}
\langle\langle({\bf v}_1-{\bf v}_2)^2
({\bf v}_3-{\bf v}_4)^2\rangle\rangle
\propto R^{-2\Delta_1}
\label{co06} \,,
\end{equation}
where ${\bf r}_1$, ${\bf r}_2$ and ${\bf r}_3$, ${\bf r}_4$ are pairs of
nearby points separated by the large distance $R$. The $r$-dependence
of the function $f^{(1)}(r)$ can also be established if one remembers that
the general scaling behavior of the correlation function of velocity
differences in the left-hand of (\ref{co06}) is determined by the
conventional KO-41 index $-4/3$. Comparing this index with the scaling
behavior (\ref{co06}) we conclude that $f^{(1)}(r)\propto
r^{\Delta_1+2/3}$.  Note that $\Delta_1+2/3=2-\Delta$ where $\Delta$ is
the anomalous dimension introduced in the preceding sections.

Besides the terms of the expansion of $({\bf v}_1-{\bf v}_2)^{2}$
over the scalar fields we should take into account also the terms of
the expansion over the vector and tensor fields. The first term of
this expansion is
\begin{equation}
({\bf v}_1-{\bf v}_2)^{2}
\to f^{(1)}_{\alpha}({\bf r}_{1}-{\bf r}_2)
O_{1,\alpha}\left(({\bf r}_1+{\bf r}_2)/2\right)
\label{co08} \,,
\end{equation}
where the vector ${\bf f}^{(1)}({\bf r})$ is directed along ${\bf r}$ and
$f^{(1)}_{\alpha}({\bf r})\propto r^{2n/3+\Delta_{1,1}}$.  Note that the
terms of the expansion of $({\bf v}_1-{\bf v}_2)^{2}$ over pseudoscalars
or over pseudovectors are absent since they are forbidden by the inversion
symmetry. The terms (\ref{co06},\ref{co08}) give us at $R\gg
r_{12},r_{34}$
\begin{eqnarray}
&& \langle\langle({\bf v}_1-{\bf v}_2)^{2}
({\bf v}_3-{\bf v}_4)^{2}\rangle\rangle
\nonumber \\
&\simeq &
f^{(1)}(r_{12})f^{(1)}(r_{34})\langle O_1({\bf R}) O_1(0)\rangle
\nonumber \\
&&+ f^{(1)}_{\alpha}(r_{12})f^{(1)}(r_{34})
\langle O_{1,\alpha}({\bf R}) O_1(0)\rangle
\nonumber \\
&&+ f^{(1)}(r_{12})f^{(1)}_{\gamma}(r_{34})
\langle O_1({\bf R}) O_{1,\gamma}(0)\rangle
\nonumber \\
&&+ f^{(1)}_{\alpha}(r_{12})f^{(1)}_{\gamma}(r_{34})
\langle O_{1,\alpha}({\bf R}) O_{1,\gamma}(0)\rangle
\label{co09} \,,
\end{eqnarray}
where {\bf R} is the vector beginning at $({\bf r}_3+{\bf r}_4)/2$ and
ending at $({\bf r}_1+{\bf r}_2)/2$; $r_{12}=|{\bf r}_1-{\bf r}_2|$ and
$r_{34}=|{\bf r}_3-{\bf r}_4|$. We see that there are three different
contributions to the correlation function which behave $\propto
R^{-2\Delta_1}$, $\propto R^{-\Delta_1-\Delta_{1,1}}$ and $\propto
R^{-2\Delta_{1,1}}$ respectively.  These three contributions can be
distinguished by their angular dependence on the angles between {\bf R}
and separation vectors ${\bf r}_1-{\bf r}_2$ and ${\bf r}_3-{\bf r}_4$.

The proposed scheme can easily be generalized for all even powers $({\bf
v}_1-{\bf v}_2)^{2n}$. We again expect that the principal term arising as
a result of ``fusion'' is
\begin{equation}
({\bf v}_1-{\bf v}_2)^{2n}
\to f^{(n)}(r_{12})O_1\left(({\bf r}_1+{\bf r}_2)/2\right)
\label{co07} \,,
\end{equation}
where $r_{12}=|{\bf r}_1-{\bf r}_2|$. The expression (\ref{co07}) leads to
the conclusion that the asymptotic behavior of any correlation function
$\langle\langle({\bf v}_1-{\bf v}_2)^{2n} ({\bf v}_3-{\bf v}_4) ^{2m}
\rangle \rangle$ is the same as (\ref{co06}).  The scaling behavior of the
function $f^{(n)}(r)$ is as follows:  $f^{(n)}(r)\propto r^{2n/3 +\Delta_1
}$.  Therefore we can establish the dependence of $\langle\langle({\bf
v}_1-{\bf v}_2)^{2n} ({\bf v}_3-{\bf v}_4)^{2m}\rangle\rangle$ not only on
large separations but also on small separations.

Then we should take into consideration the terms of the expansion of
$({\bf v}_1-{\bf v}_2)^{2n}$ over vector and tensor fields.  The principal
term of the expansion of $({\bf v}_1-{\bf v}_2)^{2n}$ over the vector
fields is as (\ref{co08}). This term produces for the correlation function
$\langle\langle({\bf v}_1-{\bf v}_2)^{2n} ({\bf v}_3-{\bf
v}_4)^{2m}\rangle\rangle$ the same structure as (\ref{co09}). In principle
one could take into account also the terms of the expansion of $({\bf v}_1
-{\bf v}_2)^{2n}$ over high-order tensorial  fields.  These terms would be
relevant if the scaling dimension $\Delta_{k,1}$ of a $k$-order tensorial
field is smaller than $\Delta_{1,1}$. In this case the angle-dependent
contribution to the correlation function $\langle\langle({\bf v}_1-{\bf
v}_2)^{m} ({\bf v}_3-{\bf v}_4)^{n} \rangle\rangle$ will be determined by
the smallest value $\Delta_{;1}$ of $\Delta_{k,1}$.  Note that
$\Delta_{;1}\le 2/3$ since $\Delta_{2,1}\le 2/3$.  The main contributions
to the correlation function $\langle\langle({\bf v}_1-{\bf v}_2)^{2n}
({\bf v}_3-{\bf v}_4)^{2m}\rangle\rangle$ in this case can be represented
like (\ref{co09}). The first contribution $\propto R^{-2\Delta_1}$ will
not depend on the angles, the second contribution $\propto
R^{-\Delta_1-\Delta_{;1}}$ is the sum of two terms depending on the angle
between {\bf R} and ${\bf r}_{1}-{\bf r}_{2}$ or on the angle between {\bf
R} and ${\bf r}_{3}-{\bf r}_{4}$ only and the last term $\propto
R^{-2\Delta_{;1}}$ depends on both angles. This is again a point where a
conformal symmetry would reveal itself : if the system possesses this
symmetry then the contribution $\propto R^{-\Delta_1-\Delta_{;1}}$ is
absent.

Above we have considered the even powers of the velocity differences.  Let
us now analyze the correlation functions of odd powers. First we consider
the special case of the first power since the difference ${\bf v}_1-{\bf
v}_2$ possesses the normal KO-41 dimension. The relation (\ref{co03})
shows that the main term of the expansion of this difference in a series
over local fields is $\nabla_\alpha v_{\beta}$. This means, for example,
that $\langle(v_{1\alpha}-v_{2\alpha})(v_{3\beta}-v_{4\beta}) \rangle
\propto R^{-4/3}$. Consider now the correlation function $\langle
(v_{1\alpha} -v_{2\alpha})({\bf v}_3-{\bf v}_4)^{2n}\rangle$.  As we have
seen the correlation function $\langle{\bf v}O_n\rangle$ is zero for any
scalar  field $O_n$.  Therefore only the term of the type (\ref{co08})
should be taken into account in the expansion of $({\bf v}_3-{\bf
v}_4)^{2n}$, giving
\begin{equation}
\langle (v_{1\alpha}-v_{2\alpha})({\bf v}_3-{\bf v}_4)^{2n}\rangle
\propto R^{-2/3-\Delta_{;1}}
\label{co10} \,,
\end{equation}
where $\Delta_{;1}$ as above is the smallest exponent of tensor fields
entering also the correlation function $\langle\langle({\bf v}_1-{\bf
v}_2)^{m} ({\bf v}_3-{\bf v}_4)^{n}\rangle\rangle$. The vector structure
of the correlation function (\ref{co10}) is determined both by the vector
{\bf R} and the vectors ${\bf r}_{1}-{\bf r}_{2}$ and ${\bf r}_{3}-{\bf
r}_{4}$. Of course among the fields $A_n$ in the expansion (\ref{co02})
for $({\bf v}_3-{\bf v}_4)^{2n}$ there is a term with $s_{\alpha\beta}$.
This means that in any case there is a term $\propto R^{-4/3}$ in the
correlation function $\langle (v_{1\alpha}-v_{2\alpha})({\bf v}_3-{\bf
v}_4)^{2n}\rangle$.  If $\Delta_{;1}<2/3$ then (\ref{co10}) decreases
slower with increasing $R$. This is again a point where conformal symmetry
could be checked: it admits only the behavior $\propto R^{-4/3}$.

Now consider a general odd power of the velocity difference
$(v_{1\alpha}-v_{2\alpha})({\bf v}_1-{\bf v}_2)^{2n}$. The first terms of
its expansion into a series over the fields $A_n$ has actually the same
structure as (\ref{co07},\ref{co08}):
\begin{eqnarray}
&& (v_{1\alpha}-v_{2\alpha})({\bf v}_1-{\bf v}_2)^{2n}\to
\nonumber \\
&& f^{(n)}_\alpha(r_{12})O_1({\bf r})+
f^{(n)}_{\alpha\beta}(r_{12})O_{1,\beta}({\bf r})
\label{co11} \,.
\end{eqnarray}
Here $f^{(n)}_\alpha(r)\propto r^{(2n+1)/3+\Delta_1}$, $f^{(n)}_{\alpha
\beta \gamma}(r) \propto r^{(2n+1)/3+\Delta_{1,1}}$.  Analogously the
terms of the expansion of $(v_{1\alpha}-v_{2\alpha})({\bf v}_1-{\bf
v}_2)^{2n}$ over tensorial fields can be introduced. We see that the terms
of the expansion of an odd power of the velocity difference are expressed
via the same fields as the expansion of even powers. Therefore the
behavior of the mutual correlation functions of the odd-odd and of the
odd-even correlation functions at large separations will be the same as
the behavior of the even-even correlation function.  Terms with different
scaling exponents can in principle be separated on the basis of their
angular dependence which for the odd powers is more complicated than for
the even ones.

\section*{Conclusion}
\label{sec:concl}

We have investigated the scaling behavior of the correlation functions of
the turbulent velocity in the inertial subrange of scales. We have argued
that there is no physical reason for the correlation functions of the
velocity differences to deviate from the behavior predicted by
Kolmogorov's dimensional estimates in the limit Re$\to \infty$.  This
conclusion is confirmed by a rigorous treatment in the framework of the
Wyld diagrammatic technique beginning with the Navier-Stokes equation:
after the Belinicher-L'vov resummation both infrared and ultraviolet
divergencies are absent for the Kolmogorov solution determined by
(\ref{veloc}). However correlation functions of powers of the velocity
derivatives possess an anomalous scaling behavior even in the limit Re$\to
\infty$. Their scaling exponents differ from those predicted by simple
KO-41 values. We present the physical picture which leads to anomalous
exponents: they appear as a result of the telescopic multi-step eddy
interaction producing anomalous contributions to the correlation
functions. This physical picture can also be reproduced on the
diagrammatic level where we have extracted a series of logarithmically
diverging diagrams whose summation leads to renormalization of the normal
KO-41 dimensions. An infinite set of anomalous dimensions should arise.
Thus the situation appears to be analogous to that in the theory of second
order phase transitions.

In order to describe the scaling behavior of correlation functions of
local fields built up from the velocity derivatives a set of fields $A_n$
with definite scaling exponents $\Delta_n$ were introduced. Any local
field can be expanded into the series over the fields $A_n$. The principal
scaling behavior of the correlation functions is determined only by the
first terms of this expansion. The symmetry classification of the fields
$A_n$ enables one to predict some relations between different correlation
functions. We have formulated also some restrictions imposed by the
incompressibility condition and propose some tests enabling the
experimental testing of conformal symmetry of the turbulent correlation
functions. We have demonstrated that anomalous scaling behavior should be
revealed in the asymptotic behavior of correlation functions of velocity
differences, and we have proposed a way to extract the anomalous exponents
from experiment.

Now some words concerning the applicability region of our theory.  We
considered the case of infinitely large $Re$ where in our consideration
the KO-41 scaling (\ref{veloc}) for the velocity correlation functions is
realized. However deviations from the KO-41 values $\zeta_n=n/3$ of the
exponents (\ref{J08}) of the structure functions (\ref{J07}) are observed
in experiment \cite{BCTB,SSJ3} and numerics \cite{BO94}. We believe that
this is related to the finite values of Re, since in our meaning a wide
region of scales down to the viscous length $\eta$ with a crossover
behavior of turbulent correlation functions should occur. Actually this
crossover may be observed in experiments. If this crossover behavior is
characterized by a single scaling then all our conclusions could also be
applied to this crossover region and only small corrections should be
introduced into the values of the KO-41 exponents noted in Section
\ref{sec:fusrl}. However if this crossover behavior is related to a
multifractal picture then our conclusions are not valid, and this case
needs a separate consideration.

\acknowledgments

Useful discussions with Itamar Procaccia and Stefan Thomae are gratefully
acknowledged. We are grateful for the support of the Landau-Weizmann
program (V.V.L.) and of the Minerva Center of Complex Physics (V.S.L.). It
is a pleasure for us to thank Adrienne Fairhall for help.

\end{document}